\pdfoutput=1

\documentclass[runningheads,a4paper]{sig-alternate}
\usepackage[table]{xcolor}
\usepackage{balance}

\usepackage{footnote}
\usepackage{amssymb}
\usepackage{hyperref}
\usepackage{url}
\usepackage{array}
\usepackage{graphicx}
\usepackage{listings}
\usepackage{multirow}
\usepackage{rotating}
\usepackage{pgfplots}
\usepackage{caption}
\usepackage{subcaption}
\usepackage{tikz}
\usetikzlibrary{backgrounds,automata,arrows,shadows,shadows.blur}

\newcolumntype{L}[1]{>{\raggedright\let\newline\\\arraybackslash\hspace{0pt}}m{#1}}
\newcolumntype{C}[1]{>{\centering\let\newline\\\arraybackslash\hspace{0pt}}m{#1}}
\newcolumntype{R}[1]{>{\raggedleft\let\newline\\\arraybackslash\hspace{0pt}}m{#1}}

\pgfplotsset{compat = 1.5}

\newenvironment{customlegend}[1][]{%
   \begingroup
    \csname pgfplots@init@cleared@structures\endcsname
    \pgfplotsset{#1}%
}{%
    \csname pgfplots@createlegend\endcsname
    \endgroup
}%

\def\addlegendimage{\csname pgfplots@addlegendimage\endcsname}

\makeatletter
\g@addto@macro\normalsize{%
  \setlength\abovedisplayskip{3pt}
  \setlength\belowdisplayskip{3pt}
  \setlength\abovedisplayshortskip{3pt}
  \setlength\belowdisplayshortskip{3pt}
}
\makeatother

\hypersetup{%
    colorlinks=true, linktocpage=true, pdfstartpage=3, pdfstartview=FitV,%
    breaklinks=true, pdfpagemode=UseNone, pageanchor=true, pdfpagemode=UseOutlines,%
    plainpages=false, bookmarksnumbered, bookmarksopen=true, bookmarksopenlevel=1,%
    hypertexnames=true, pdfhighlight=/O,
    urlcolor=black, linkcolor=black, citecolor=black, 
} 

\begin{document}

\title{MobileAppScrutinator:\\A Simple yet Efficient Dynamic Analysis Approach for Detecting Privacy Leaks across Mobile OSs}
 \numberofauthors{2}
 \author{
        \alignauthor 
        Jagdish Prasad Achara\\ Vincent Roca and Claude Castelluccia \\
        \affaddr{Inria, Grenoble, France}\\
        \email{firstname.lastname@inria.fr}
        \alignauthor
        Aur\'elien Francillon \\
        \affaddr{Eurecom, Sophia-Antipolis, France}\\
        \email{aurelien.francillon@eurecom.fr}
 }
\maketitle

\begin{abstract}
Smartphones, the devices we carry everywhere with us, are being heavily tracked 
and have undoubtedly become a major threat to our privacy. 
As ``Tracking the trackers'' has become a necessity, various static and dynamic 
analysis tools have been developed in the past. 
However, today, we still lack suitable tools to detect, measure and compare the 
ongoing tracking across mobile OSs. 
To this end, we propose \emph{MobileAppScrutinator}, based on a 
\emph{simple yet efficient} dynamic analysis approach. 
To demonstrate the current trend in tracking, we select 140 most representative 
apps available on both Android and iOS AppStores and test them with 
MobileAppScrutinator. 
In fact, choosing the same set of apps on both Android and iOS also enables us 
to \emph{compare the ongoing tracking} on these two OSs.
Finally, we also discuss the effectiveness of privacy safeguards available on 
Android and iOS. 
We show that neither Android nor iOS privacy safeguards in their present state 
are completely satisfying.

\end{abstract}

\section{Introduction}

Smartphones no longer involve only the user and the communication service 
(GSM/CDMA) provider.
The revolution of the AppStore model for application distribution brings a large 
number of new actors.
In the literature, service providers to whom the user directly interacts with 
are considered as first-party, the user being the second-party.
However, there are many additional actors whose services are not directly used 
by the end user, and whose presence is not obvious to most users, are called 
the third-parties.
Third parties are, for example, Advertisers and Analytics (A\&A) companies, 
application performance monitors, crash reporters, or push notification senders.
The situation has become even more complex with the development of new 
advertisement models
like Mobile Ad Networks or Ad exchange networks for real-time bidding (RTB).

Depending on the service provided, a user may accept to exchange data with a 
first-party, in general following legal terms and conditions upon which they 
mutually agree.
However, the data collection by third-parties without explicit user consent is 
more problematic. 
Due to economic reasons, the A\&A companies are the dominant third parties and 
the most privacy intrusive.
Indeed, in order to increase their revenue, advertisers want to send 
personalized Ads to the user. 
Therefore A\&A companies are incited to collect as much information as possible 
to better profile user's interests and behavior. 
In order to achieve this goal, they need a way to identify the smartphone/user 
via an identifier that can uniquely be associated with a smartphone/user. 
This whole process of data collection is called 
{\bf ``third-party smartphone tracking"} or simply {\bf ``tracking"} and the 
process of showing user-specific Ads based on user profile 
is called {\bf ``targeted advertising"}. 

Smartphone tracking and targeted-advertising are acceptable if the user is 
aware of it and if he agrees to receive targeted Ads based on his personal 
interests. 
Some users could also find the presence of third-parties on the smartphone 
beneficial.
However, problems arise when A\&A companies collect Personally-Identifiable 
Information (PII) without users' knowledge. 
In fact, Ad libraries sometimes also include APIs through which an application 
can deliberately leak user PII~\cite{DBLP:journals/corr/BookW13}. 
This creates serious privacy risks for users. 
With the rapidly growing number of smartphones, people are increasingly exposed 
to such risks. 
Moreover, a smartphone is particularly intrusive, revealing all user movements 
as it is equipped with many sensors, and it stores a plethora of information 
either generated by these sensors, by the telephony services (calls and SMS), 
or by the user himself (e.g., calendar events and reminders). 
Finally, various scandals in the past 
(e.g.,\cite{url:WSJ_WhatTheyKnowSeries_Mobile, url:Twitter_Path}) make it 
difficult to trust all these actors present on smartphones.

\paragraph{Motivation}
As ``Tracking these trackers" has become a necessity, various tools have been 
developed in the past. 
These tools are based on either static analysis or dynamic analysis or 
interception of network traffic. 
Even though static analysis techniques scale well, they generally fail on 
obfuscated applications and therefore, are not suited to accurately detect and 
measure the ongoing tracking.
Similarly, network interception is often unable to handle SSL traffic.
Dynamic analysis techniques for detection and measurement of PII leaks are 
available on Android but not on iOS. 
As we lack suitable dynamic analysis tools readily available on 
\emph{both} Android and iOS, there is no measurement study in the literature 
which provides concrete evidence of ongoing tracking as well as the comparision 
across mobile OSs.

\paragraph{Contributions}
The contributions of this paper are threefold:
\begin{enumerate} 
    \item We present our dynamic analysis platform, \emph{MobileAppScrutinator}, 
          which detects and measure the ongoing tracking on Android and iOS.
          It is the first dynamic analysis platform for iOS to detect private 
      data leakage.
	
    \item We detect private information leaks by applications over the Internet, 
          when they leak it in clear-text or over SSL. MobileScrutinator 
      detects the PII leakage even if the App obfuscates the data (by hashing 
      or encrypting) before sending over the network in clear-text or 
      through SSL.
     Detection of modified PII is a key for reliable measurements as some 
     identifiers (e.g., WiFi MAC address, AndroidID, IMEI) are often modified 
     before being sent. 

    \item We test 140 popular applications with \emph{MobileAppScrutinator}, on 
    both Android and iOS, and report our findings with a comparision of 
    ongoing tracking on these two platforms.
    
    \item Finally, we discuss the effectiveness of privacy safeguards available 
    on both Android and iOS. 
    We show that neither Android nor iOS privacy safeguards in their present 
    state are completely satisfying.
    
\end{enumerate}

\section{Related work}
\label{sec:Related_Work}

Our work can be compared with
existing works on two axes: 1) Tracking measurement technology/tools and 2)
Measurement of PII leakage. 
Below we discuss and compare our work with some most representative works along 
these
two axes.

\subsection{Tracking measurement technology}

Tools to measure the ongoing tracking might be based on either interception of
generated network traffic, or static analysis of the application code, or 
dynamic analysis of applications.

\paragraph{Interception of generated network traffic} This approach is based on
snooping the network data using Man-In-The-Middle (MITM) proxy.  For example,
MobileScope~\cite{mobilescope}, based on MITM proxy, was used in WSJ 
study~\cite{url:WSJ_WhatTheyKnowSeries_Mobile} to
investigate the top 100 applications on both Android and iOS. However, this
technique cannot be used to intercept the SSL traffic which seriously
limits the effectiveness of this approach; as we see in 
Sections~\ref{sec:In_App_Tracking} and~\ref{sec:PI_Collection} that almost half 
of PII leakage is through SSL.
Additionally, MITM approach will not be able to detect reliably the leakage of 
PII generated by the system (values not known to the user and therefore, could 
not be searched in the network traffic). 
This includes different PII, for example, unique IDs generated and shared by 
applications and user location. 
Also, MITM based approach would fail in cases where user PII is modified 
(e.g., hashed or encrypted) before being sent (and, as we will show, this is a 
rather common practice). 
Finally, being a network packet analysis approach, it is not always easy or 
feasible to identify the application having generated the monitored traffic, 
which makes the (manual) analysis even more complex. 
MobileAppScrutinator, in contrast, makes analysis directly at the operating 
system level, and thus does not suffer from such limitations.

\paragraph{Static analysis} 
Past works (PiOS ~\cite{pios} on iOS, FlowDroid~\cite{arzt2014flowdroid}, 
ScanDroid~\cite{fuchs2009scandroid}, CHEX~\cite{lu2012chex}, 
AndroidLeaks~\cite{gibler2012androidleaks}, SymDroid~\cite{jeon2012symdroid}, 
ScanDal~\cite{kim2012scandal} and AppIntent~\cite{yang2013appintent} on Android) 
are based on static analysis to detecte a flow of data from a PII source to a 
network sink. 
These works can be classified in two categories: 
1) static tainting-based (e.g., FlowDroid~\cite{arzt2014flowdroid}, 
ScanDroid~\cite{fuchs2009scandroid}, CHEX~\cite{lu2012chex}, 
AndroidLeaks~\cite{gibler2012androidleaks}) and 2) symbolic execution based 
(e.g., SymDroid~\cite{jeon2012symdroid}, ScanDal~\cite{kim2012scandal} and 
AppIntent~\cite{yang2013appintent}). 
Among the ones based on symbolic execution, SymDroid~\cite{jeon2012symdroid} 
designs a symbolic executor based on their simple version of Dalvik VM, i.e., 
micro-dalvik. 
Similarly, ScanDal~\cite{kim2012scandal} designs an intermediate language, 
called Dalvik Core, and collects all the program states during the execution of 
the program for all inputs. 
Considering the Android's special event-driven paradigm, 
AppIntent~\cite{yang2013appintent} proposes a more efficient event-space 
constraint guided symbolic execution. 
On iOS, PiOS was designed for binaries compiled with GCC/G++ compiler and since
then, Apple switched to LLVM compiler. 
Therefore, PiOS needs to be adapted to support the analysis of binaries compiled 
with LLVM. Furthermore, PiOS is not available publicly, so one needs to build it 
from scratch to use it to detect and measure the ongoing tracking.
In general, static analysis techniques do scale well but they lack dynamic 
information tracking and therefore, lead to false negatives.

\paragraph{Dynamic analysis} 
TaintDroid~\cite{taintdroid} and PMP~\cite{agarwal2013protectmyprivacy} are 
based on dynamic analysis on Android and iOS respectively.
On Android, TaintDroid is a dynamic taint-based technique to detect and measure 
private data leakage. However, TaintDroid has its own limitations: 
1) taint-based tracking can be easily circumvented using indirect information 
flows ~\cite{sarwar2013effectiveness} 2) requires to make a trade-off between 
false positives and false negatives (\cite{taintdroid} did not taint IMSI due 
to false positives) 3) misses native code (both for taint propagation and as a 
source of information).
On iOS, PMP~\cite{agarwal2013protectmyprivacy} is a dynamic/runtime tool that 
offers
the functionality of choosing access to what information a user is willing to 
share with a particular application. 
As iOS's own privacy control feature provides the same functionality, 
PMP~\cite{agarwal2013protectmyprivacy} is an enhancement in terms of the number 
of different types of private data considered. In fact, PMP fails to notify 
users if the accessed information is being sent to a remote server or not.
As existing dynamic analysis tools on iOS are not sufficient to measure the 
ongoing tracking on iOS, one possibility could have been a taint-based dynamic 
analysis technique. 
However, this is not possible to do on iOS device because the code of 
applications is native (C, C++ and Objective-C). 
Propagating taint would require to emulate native code, which involves serious 
changes to the system and would have a significant performance penalty. 
Therefore, we opted for a dynamic analysis approach, 
described in the next section, which can be used on both Android and iOS. 
To measure the effectiveness of MobileAppScrutinator on Android with TaintDroid, 
we perform tests on identical set of applications using TaintDroid and 
MobileAppScrutinator. 
We found that MobileAppScrutinator could detect a lot more PII leakage than 
TaintDroid.



\subsection{Measurement of PII leakage}
To best of our knowledge, no previous work provides a complete
picture of tracking on Android and iOS.
Web-browser tracking has been thoroughly
studied~\cite{Mayer:2012,Roesner:2012}, but it is not the case with smartphone
tracking.
We are first to provide detailed analysis and measurement data for both 
Android and iOS.
\cite{han2011study} sheds some light
on third-party tracking being taken place on Android using TaintDroid but is
not as comprehensive as ours. 
Also, all other static and dynamic analysis tools proposed in the literature, 
for example, PiOS~\cite{pios} and TaintDroid~\cite{taintdroid}, analyzed some 
applications and presented a number of applications leaking user PII, but none 
of them presented a complete analysis as we do in 
Section~\ref{sec:In_App_Tracking} and~\ref{sec:PI_Collection}. 
Also, as tracking technologies rapidly change with OS revisions, it is crucial 
to have up-to-date tools and a recent picture of tracking technology. 
Furthermore, we also consider the remote servers where PII is sent and 
attempted to distinguish them among first and third-parties, with the 
available information. 

\section{MobileAppScrutinator}
\label{sec:MobileAppScrutinator}
\paragraph{Design choices}
From tracking detection and measurement point of view, it is ideal to analyse 
what applications are doing at operating system level. 
However, we want to have a system that does not require too intrusive 
modifications of the OSs and does not have too many false positives 
(unlike dynamic tainting-based techniques). The system should work well with 
non-malicious application on real devices so that it can be used by anyone.
Thus the design of MobileAppScrutinator starts with a simple approach: 
\emph{intercepting the source, sink and data manipulation system APIs}.
As the same approach is applied to both Android and iOS, it enables us to 
compare the tracking across mobile OSs.
Even though MobileAppScrutinator is based on this same simple approach on both 
Android and iOS, i.e., the basic governing philosophy remains the same, its 
implementation \emph{differs significantly} due to the 
differences of these OSs.

\paragraph{Overall architecture}
As developer APIs are public on both Android and iOS, we are able to identify 
all source and sink methods (i.e., methods related to access or modification of 
private data along with network operations either in clear-text or encrypted).
MobileAppScrutinator hooks these APIs and includes extra code to these APIs. 
This added extra code collects various information from the application 
environment.
Specifically, it collects information about PII being accessed, modified, or 
transmitted by an application along with the information about that application.  
Any access, or modification, or transmission of PII corresponds to an event and 
is stored locally in an SQLite database.
This database is later analyzed automatically to detect and measure privacy 
leaks.

In the next two subsections, we give details of how MobileAppScrutinator is 
implemented on Android and iOS.

\subsection{Implementation on Android}

\begin{figure}[!t]
\centering
\scalebox{0.60}{\input{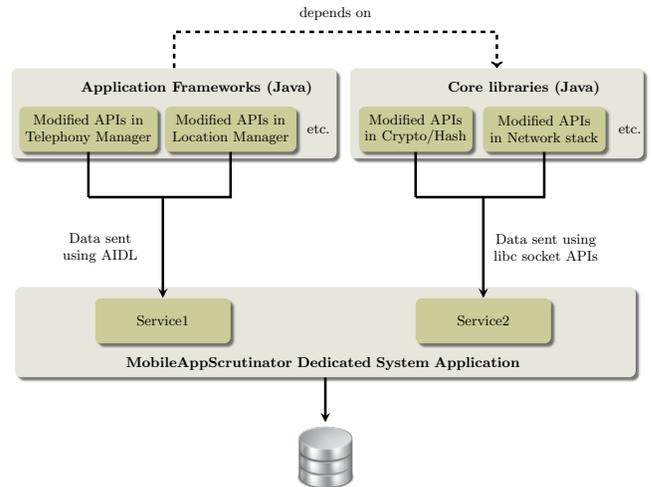}}
\caption{MobileAppScrutinator Android implementation.}
\label{fig:AndroidRealization}
\end{figure}
As the Android source (from Android Open Source Project (AOSP)) is publicly 
available, MobileAppScrutinator directly modifies source code of various APIs 
in Java frameworks. 
This modified source is compiled and a new system image is generated.
We develop and add a system application to this new system image.
This system application runs two Android services that are responsible for 
receiving data from different sources. 
The App is also responsible for storage of data in a local SQLite database.

As described earlier, our added extra code must send data to 
MobileAppScrutinator system app.
To do so, we use two methods: 
AIDL\footnote{\url{http://developer.android.com/guide/components/aidl.html}} and 
socket APIs of libc library.
Our extra coded added in Android application frameworks APIs uses AIDL to send 
data to MobileAppScrutinator system app whereas socket APIs of libc library are 
used in modified core Java frameworks. 
Here it is to be noted that Android application frameworks are written utilizing 
core Java frameworks, i.e., 
during compilation of Android source, core Java frameworks are compiled before 
the Android application frameworks.
As AIDL 
is part of the Android application framework, 
the code added (by MobileAppScrutinator) in modified core Java frameworks cannot 
use AIDL to send data to MobileAppScrutinator system application.
Therefore, to send data to MobileAppScrutinator system apps from modified APIs 
of core Java frameworks, MobileAppScrutinator uses socket APIs of libc library.
 to send data to a dedicated service running inside MobileAppScrutinator system 
 application. 
Fig.~\ref{fig:AndroidRealization} provides a broad picture of how 
MobileAppScrutinator is implemented on Android OS.

\subsection{Implementation on iOS}

\begin{figure}  
	\centering
	\begin{subfigure}[]{0.48\textwidth}
		\centering
		\scalebox{0.60}{\input{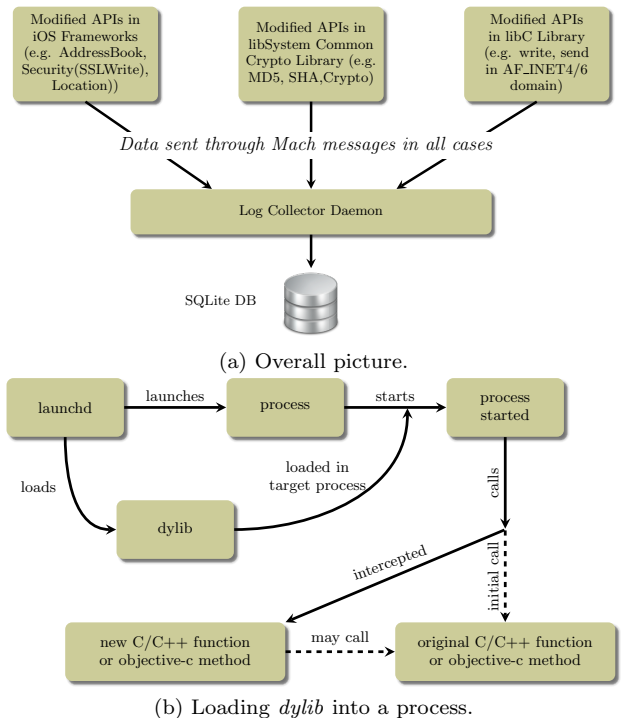}}
		\caption{Overall picture.}
		\label{fig:iOSRealization}
	\end{subfigure}
	\begin{subfigure}[]{0.48\textwidth}
		\centering
		\scalebox{0.65}{\begin{tikzpicture}

\definecolor{col1}{RGB}{102, 153, 102}
\definecolor{col2}{RGB}{51, 102, 153}
\definecolor{col3}{RGB}{124, 120, 106}
\definecolor{col4}{RGB}{230, 230, 220}
\definecolor{col5}{RGB}{204, 204, 153}
\definecolor{col6}{RGB}{204, 152, 0}
\definecolor{col7}{RGB}{180, 30, 47} 

\tikzstyle{commonStyle}=[draw=none,text centered,drop shadow,blur shadow={shadow blur steps=5},rounded corners]

\draw (0,0) node[text width = 2cm,minimum height=1.2 cm,minimum width=2.4cm,fill=col5,commonStyle] (L) {launchd};
\draw (4.5,0) node[text width = 2cm,minimum height=1.2 cm,minimum width=2.4cm,fill=col5,commonStyle] (P) {process};
\draw (9,0) node[text width = 2cm,minimum height=1.2 cm,minimum width=2.4cm,fill=col5,commonStyle] (PS) {process started};

\draw (2.25,-2.5) node[text width = 2cm,minimum height=1.2 cm,minimum width=2.4cm,fill=col5,commonStyle] (DY) {dylib};
\draw (2.25,-5) node[text width = 4cm,minimum height=1.2 cm,minimum width=4.5cm,fill=col5,commonStyle] (NW) {new C/C++ function or 
objective-c method};

\draw (9,-5) node[text width = 4cm,minimum height=1.2 cm,minimum width=4.5cm,fill=col5,commonStyle] (ORG) {original C/C++ function or objective-c method};

\draw (9,-2.5) node[outer sep=0,inner sep=0] (emptyNode) {};
\draw (7,0) node[outer sep=0,inner sep=0] (emptyNode2) {};

\draw[->,ultra thick,black,>=stealth] (L) -- node[midway,above] {launches} (P);
\draw[->,ultra thick,black,>=stealth] (P) -- node[midway,above] {starts} (PS);
\draw	  (L.south) edge[->,ultra thick,black,>=stealth,out=270,in=180]  node[midway,above,xshift=-0.75cm] {loads} (DY.west);
\draw	[->,ultra thick,black,>=stealth] (PS.south) -- node[midway,above,rotate=90] {calls} (emptyNode);
\draw	[->,ultra thick,black,>=stealth] (emptyNode) -- node[midway,above,rotate=20] {intercepted} (NW.15);
\draw	[->,ultra thick,dashed,black,>=stealth] (NW.east) -- node[midway,above] {may call} (ORG.west);

\draw	[->,ultra thick,dashed,black,>=stealth] (emptyNode) -- node[midway,above,rotate=90] {initial call} (ORG.north);

\draw	 (DY.east) edge [->,ultra thick,black,>=stealth,out=0,in=270] node[midway,above,text width=2cm,align=center,xshift=-.7cm,yshift=-0cm] {loaded in target process} (emptyNode2);

\end{tikzpicture}}
		\caption{Loading \emph{dylib} into a process.}
		\label{fig:DyLib_Loading_Process_iOS}
	\end{subfigure}
	\caption{MobileAppScrutinator iOS implementation.}
	\label{fig:iOS_AppScrutinator_Realization}
\end{figure}

Fig.~\ref{fig:iOSRealization} depicts an overall picture of implementation of
MobileAppScrutinator on iOS. 
The APIs of interest in iOS frameworks are modified to capture and send data 
related to application context as well as the PII being accessed, modified, 
or transmitted (clear-text or SSL). 
The data communication between various processes running our custom code and 
the daemon is through \emph{mach} messages.
In order to execute self-signed code and get privileged access to the
system, the default iOS software stack needs to be modified to remove the
restrictions imposed by Apple (a technique known as ``Jailbreaking" in the iOS
world).

On iOS, developers may write code in C, C++ and Objective-C languages.
In fact, all iOS executables are linked with the Objective-C 
runtime~\cite{obj-c-runtime} and this runtime environment provides a method 
called  \emph{method\_setImplementation}. 
We use this method to change the implementation of existing Objective-C methods 
whereas to change the implementation of C/C++ functions, we use the trampoline 
technique~\cite{trampoline}.
MobileSubstrate~\cite{mobilesubstrate}, an open-source framework, greatly 
simplifies this task. 
Finally, the source code responsible for modification of various APIs of 
interest is compiled into a dynamic library (\emph{dylib}) which is loaded 
using \emph{launchd}~\cite{launchd}, into all or a subset of running processes.
Fig.~\ref{fig:DyLib_Loading_Process_iOS} depicts how a \emph{dylib} is loaded 
into a process using \emph{launchd}.

\subsection{Post-analysis of SQLite Data}

The events stored in local SQLite database are processed by automated Python 
scripts.
It is a two-step process: a first pass over the database on a per-application 
basis results
into a JSON file, and a second pass over the JSON file derives various
statistics.

Our first level analysis consists of the following steps:
\begin{enumerate}
		
	\item Find all types of PII accessed by each application.

	\item Check if PII is really sent over the network or not, and if yes, 
    to which
		server it is sent to. 

	\item Search for the PII in the input to data modification APIs 
    (cryptographic and hashing) and
		if found, look for the result in the data sent over the network.

\end{enumerate}

First-level analysis results into a JSON file that stores 1)
accessed PII, 2) PII passed to encryption or hash APIs 3)
(un)modified PII sent over the Internet in cleartext or using SSL.
Once the first pass over the database is finished, the resulted JSON file 
containing
per-App details is processed to infer or derive various statistics. Here it is
worth to mention that various PII accessed by an application are searched only 
in the network traffic and hash/cryptographic calls of that specific application. 

In various APIs, the access to data is at byte level and
in this case, the raw bytes are first attempted to be decoded using UTF-8
encoding. Since a different encoding may be used or in case of binary data, the 
hexadecimal
representation of these raw bytes is also stored alongside. 
Searching in the
network, or in the input to cryptographic or hash APIs is done for both UTF-8
encoded data and hexadecimal representation of the raw bytes. 

\subsection{Limitations}
The PII leakage would remain undetected if the data is modified by the 
application developer using
custom data modification functions before sending it over the network. 
If the PII
is modified using OS provided data modification APIs (e.g., encryption, hashing) 
before sending over
network, MobileAppScrutinator would correctly be able to detect the PII leakage.
As an App developer is not bound to use system provided hash and
encryption APIs, MobileAppScrutinator might miss some PII leakage instances.

In addition to this, the current AppScrutinator implementation on Android only 
supports the Java code.
Therefore it currently does not detect PII leakage that involves calls to C/C++ 
APIs using the JNI framework. 
However, this is not a limitation of the approach and MobileAppScrutinator could 
easily be extended to handle such cases.
handle JNI calls.

\section{Experimental Setup}
\label{sec:experimental_setup}
In order to investigate the tracking mechanisms being used by
third-parties, we test 140 representative
 free Apps available both on Android and iOS (most popular Apps in each category).  Experiments have been conducted on iOS 6.1.2 and
Android 4.1.1\_r6.

We manually ran applications for approximately one minute each. We could interact with some applications during this one minute duration as others required the user to log in or sign up.
We did not sign up or log in as our ultimate goal
was not to track the manually entered user PII 
but the seamless background tracking done without any user intervention/interaction.
Also, we did not try to cover all possible execution paths, indeed third-party library code generally starts execution when the application is first launched.

Apart from device or operating system unique identifiers and information, we 
also entered other synthetic information such as addressbook, calendar events, accounts 
etc. This enables us to know if such data is accessed and transmitted by apps.

\section{Cross-App Third-Party Tracking} 
\label{sec:In_App_Tracking}

Smartphone users mostly use dedicated apps rather than websites for accessing 
services, essentially because of the relatively small screen size and the lack 
of mobile-optimized web pages (even if this later aspect has largely improved).
Therefore user tracking is no longer performed only through 
``third-party cookies'' in web browsers but also in apps through dedicated 
identifiers. In this section, we shed some light on this ongoing user tracking 
in apps through stable identifiers.

\subsection{Unique identifiers from the system}

First of all, let us consider the system level unique identifiers.
The situation is rather different depending on the target OS.

\subsubsection{Android.}

Various system identifiers are available on Android.
A user permission is required to access hardware-tied identifiers (IMEI and 
Wi-Fi MAC address) as well as SIM-tied identifiers (IMSI and phone number). 
OS-tied identifiers (Serial Number and AndroidID) identifiers are freely 
available to be accessed.

\begin{figure}[t]  
\centering
	\begin{subfigure}[]{0.75\linewidth}
		\scalebox{0.75}{\begin{tikzpicture}
\definecolor{Col1}{HTML}{CC0000}
\definecolor{Col2}{HTML}{CCCC99}
\definecolor{Col3}{HTML}{003366}
\definecolor{Col4}{HTML}{996600}
\definecolor{Col6}{HTML}{669966}
\definecolor{Col7}{HTML}{666699}
\definecolor{Col8}{HTML}{FFCC00}
\begin{axis}[
ybar stacked,
ymajorgrids,
ymin=0,
axis line style={draw=none},
y tick label style={major tick length=0pt},
x tick label style={major tick length=0pt},
myplotShadow/.style={drop shadow={shadow yshift=0.25pt, shadow xshift=0.65pt}},
myplot/.style={draw=none,area legend,draw opacity=0,myplotShadow},
enlargelimits=0.05,
legend style={at={(0.7,0.95)},
anchor=north,legend columns=-1},
ylabel={\#Apps},
symbolic x coords={Android ID,IMEI,WiFi MAC,Serial No,IMSI,PhoneNumber},
xtick=data,
nodes near coords,
nodes near coords align={vertical},
x tick label style={rotate=45,anchor=east},
ymin=0,ymax=65,
]
\addplot[myplot,fill=Col3] coordinates {(Android ID,28)(IMEI,20)(WiFi MAC,9)(Serial No,7)(IMSI,5)(PhoneNumber,3)};
\addplot[myplot,fill=Col2] coordinates {(Android ID,13)(IMEI,8)(WiFi MAC,2)(Serial No,3)(IMSI,0)(PhoneNumber,0)};
\end{axis}
\end{tikzpicture}}
		\caption{Android.\label{fig:Android_NoOfApps}}
		
	\end{subfigure}
	\begin{subfigure}[]{0.75\linewidth}
		\scalebox{0.75}{\begin{tikzpicture}
\definecolor{Col1}{HTML}{CC0000}
\definecolor{Col2}{HTML}{CCCC99}
\definecolor{Col3}{HTML}{003366}
\definecolor{Col4}{HTML}{996600}
\definecolor{Col6}{HTML}{669966}
\definecolor{Col7}{HTML}{666699}
\definecolor{Col8}{HTML}{FFCC00}

\begin{axis}[
ybar stacked,
ymajorgrids,
ymin=0,
axis line style={draw=none},
y tick label style={major tick length=0pt},
x tick label style={major tick length=0pt},
myplotShadow/.style={drop shadow={shadow yshift=0.25pt, shadow xshift=0.65pt}},
myplot/.style={draw=none,area legend,draw opacity=0,myplotShadow},
ylabel={\#Apps},
symbolic x coords={Pasteboard ID,AdIdentifier,WiFi MAC,UDID,Device Name},
xtick=data,
nodes near coords,
nodes near coords align={vertical},
x tick label style={rotate=45,anchor=east},
ymin=0,ymax=65,
]
\addplot[myplot,fill=Col3] coordinates {(Pasteboard ID,63)(AdIdentifier,58)(WiFi MAC,12)(UDID,7)(Device Name,4)};
\addplot[myplot,fill=Col2] coordinates {(Pasteboard ID,0)(AdIdentifier,0)(WiFi MAC,34)(UDID,0)(Device Name,0)};
\end{axis}
\end{tikzpicture}}
		\caption{iOS.\label{fig:iOS_NoOfApps}}
	\end{subfigure}\\
	\scalebox{0.75}{\definecolor{Col1}{HTML}{CC0000}
\definecolor{Col2}{HTML}{CCCC99}
\definecolor{Col3}{HTML}{003366}
\definecolor{Col4}{HTML}{996600}
\definecolor{Col6}{HTML}{669966}
\definecolor{Col7}{HTML}{666699}
\definecolor{Col8}{HTML}{FFCC00}
\begin{tikzpicture}
    \begin{customlegend}[legend columns=5,legend style={nodes={right},draw=none,font=\small},legend entries={Unmodified, Modified}]
        \addlegendimage{draw=none,area legend,draw opacity=0,blur shadow={shadow blur radius=0.25pt,shadow yshift=-0.5pt, shadow xshift=0.5pt},fill=Col3}
        \addlegendimage{draw=none,area legend,draw opacity=0,blur shadow={shadow blur radius=0.25pt,shadow yshift=-0.5pt, shadow xshift=0.5pt},fill=Col2}
        \addlegendimage{only marks, mark=square*,draw opacity=0,fill=predictedCol,legend image post style={xshift=0.2cm}}
    \end{customlegend}
\end{tikzpicture}}
	\caption{\# Apps sending System identifiers (out of 140 apps).\label{fig:iOS_Android_NoOfApps}}
\end{figure}
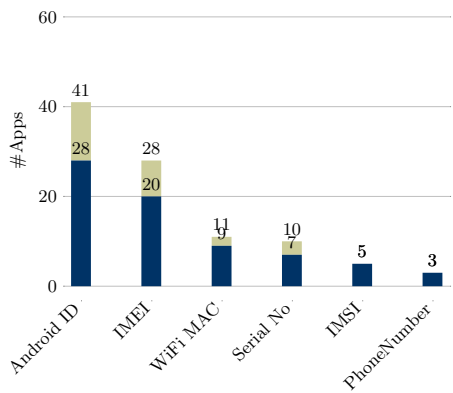
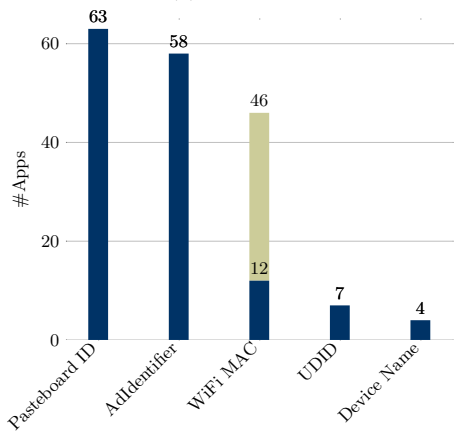

Fig.~\ref{fig:Android_NoOfApps} presents the number of android apps (out of a 
total of 140 most popular apps tested) that transmit 
various unique system identifiers over the Internet. 
We note that a significant number of apps transmit hardware-tied identifiers 
such as IMEI number and WiFi MAC address of the phone.
IMEI number alone is transmitted by 28 apps, i.e., by 20 \% apps.
This is very critical in terms of user privacy because users cannot reset 
hardware-tied identifiers (unlike a cookie in a  web browser).
Android ID is the most frequently  OS-tied identifier transmitted (by around 30\% apps).
In general, SIM-tied identifiers are the least frequently transmitted: 
IMSI is transmitted by five apps whereas three apps transmitted the phone 
number.
Interestingly, we also note that most of these identifers are transmitted 
unmodified, i.e., without hashing or encrypting, over the Internet.
This clearly demonstrates that app developers do not care at all about user 
privacy and the legislation is not hard enough to force the app developers to 
care about user privacy.

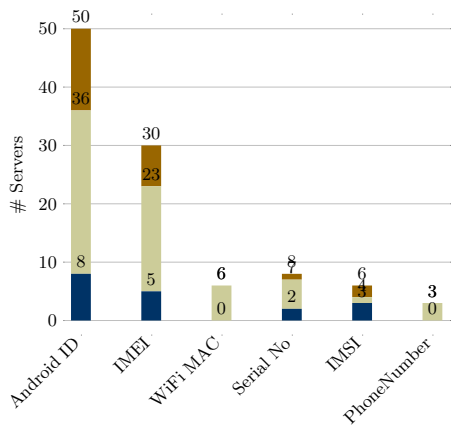
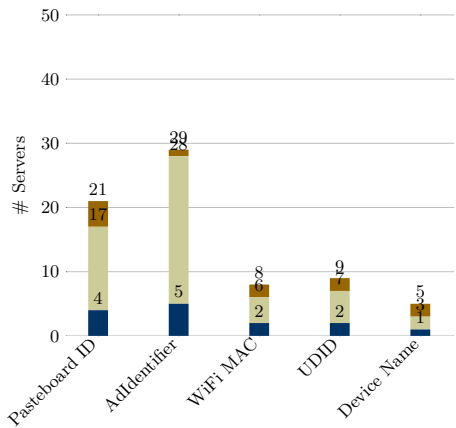
\begin{figure}[t]  
\centering
	\begin{subfigure}[]{0.75\linewidth}
		\scalebox{0.75}{\begin{tikzpicture}
\definecolor{Col1}{HTML}{CC0000}
\definecolor{Col2}{HTML}{CCCC99}
\definecolor{Col3}{HTML}{003366}
\definecolor{Col4}{HTML}{996600}
\definecolor{Col6}{HTML}{669966}
\definecolor{Col7}{HTML}{666699}
\definecolor{Col8}{HTML}{FFCC00}
\begin{axis}[
ybar stacked,
ymajorgrids,
ymin=0,
axis line style={draw=none},
y tick label style={major tick length=0pt},
x tick label style={major tick length=0pt},
myplotShadow/.style={drop shadow={shadow yshift=0.25pt, shadow xshift=0.65pt}},
myplot/.style={draw=none,area legend,draw opacity=0,myplotShadow},
enlargelimits=0.05,
legend style={at={(0.7,0.95)},
anchor=north,legend columns=-1},
ylabel={\# Servers},
symbolic x coords={Android ID,IMEI,WiFi MAC,Serial No,IMSI,PhoneNumber},
xtick=data,
nodes near coords,
nodes near coords align={vertical},
x tick label style={rotate=45,anchor=east},
ymin=0,ymax=50,
]
\addplot[myplot,fill=Col3] coordinates {(Android ID,8)(IMEI,5)(WiFi MAC,0)(Serial No,2)(IMSI,3)(PhoneNumber,0)};
\addplot[myplot,fill=Col2] coordinates {(Android ID,28)(IMEI,18)(WiFi MAC,6)(Serial No,5)(IMSI,1)(PhoneNumber,3)};
\addplot[myplot,fill=Col4] coordinates {(Android ID,14)(IMEI,7)(WiFi MAC,0)(Serial No,1)(IMSI,2)(PhoneNumber,0)};
\end{axis}
\end{tikzpicture}}
		\caption{Android.\label{fig:Android_sysIdTransmitted}}		
	\end{subfigure}
	\begin{subfigure}[]{0.75\linewidth}
		\scalebox{0.75}{\begin{tikzpicture}
\definecolor{Col1}{HTML}{CC0000}
\definecolor{Col2}{HTML}{CCCC99}
\definecolor{Col3}{HTML}{003366}
\definecolor{Col4}{HTML}{996600}
\definecolor{Col6}{HTML}{669966}
\definecolor{Col7}{HTML}{666699}
\definecolor{Col8}{HTML}{FFCC00}

\begin{axis}[
ybar stacked,
ymajorgrids,
ymin=0,
axis line style={draw=none},
y tick label style={major tick length=0pt},
x tick label style={major tick length=0pt},
myplotShadow/.style={drop shadow={shadow yshift=0.25pt, shadow xshift=0.65pt}},
myplot/.style={draw=none,area legend,draw opacity=0,myplotShadow},
ylabel={\# Servers},
symbolic x coords={Pasteboard ID,AdIdentifier,WiFi MAC,UDID,Device Name},
xtick=data,
nodes near coords,
nodes near coords align={vertical},
x tick label style={rotate=45,anchor=east},
ymin=0,ymax=50,
]
\addplot[myplot,fill=Col3] coordinates {(Pasteboard ID,4)(AdIdentifier,5)(WiFi MAC,2)(UDID,2)(Device Name,1)};
\addplot[myplot,fill=Col2] coordinates {(Pasteboard ID,13)(AdIdentifier,23)(WiFi MAC,4)(UDID,5)(Device Name,2)};
\addplot[myplot,fill=Col4] coordinates {(Pasteboard ID,4)(AdIdentifier,1)(WiFi MAC,2)(UDID,2)(Device Name,2)};
\end{axis}
\end{tikzpicture}}
		\caption{iOS.\label{fig:iOS_sysIdTransmitted}}		
	\end{subfigure}\\
	\scalebox{0.75}{\definecolor{Col1}{HTML}{CC0000}
\definecolor{Col2}{HTML}{CCCC99}
\definecolor{Col3}{HTML}{003366}
\definecolor{Col4}{HTML}{996600}
\definecolor{Col6}{HTML}{669966}
\definecolor{Col7}{HTML}{666699}
\definecolor{Col8}{HTML}{FFCC00}
\begin{tikzpicture}
    \begin{customlegend}[legend columns=5,legend style={nodes={right},draw=none,font=\small},legend entries={First-party, Third-party, Unidentified}]
        \addlegendimage{draw=none,area legend,draw opacity=0,blur shadow={shadow blur radius=0.25pt,shadow yshift=-0.5pt, shadow xshift=0.5pt},fill=Col3}
        \addlegendimage{draw=none,area legend,draw opacity=0,blur shadow={shadow blur radius=0.25pt,shadow yshift=-0.5pt, shadow xshift=0.5pt},fill=Col2}
        \addlegendimage{draw=none,area legend,draw opacity=0,blur shadow={shadow blur radius=0.25pt,shadow yshift=-0.5pt, shadow xshift=0.5pt},fill=Col4}
        \addlegendimage{only marks, mark=square*,draw opacity=0,fill=predictedCol,legend image post style={xshift=0.2cm}}
    \end{customlegend}
\end{tikzpicture}}
	\caption{\# servers where system identifiers are sent by 140 apps.	
          \label{fig:iOS_Android_ids_transmitted}
}
\end{figure}

Fig.~\ref{fig:Android_sysIdTransmitted} presents the number of servers where 
unique system identifiers are transmitted by 140 Android apps. 
The servers are classified as either first-party or third-party or 
unidentified.
First-party servers are those where domain name conincides with the app name, 
third-party servers are domains of well known advertisers and trackers, and unidentified 
are the ones where domain name is either of content providers or ip addresses 
for which no hostname information is found.
We note here that various unique system identifiers are transmitted to both 
first and third-parties. 
We also find that various unique identifiers are also sent to some IP addresses 
without any hostname information.
It is not easy to identify to whom such machines belong to and why these 
identifiers are transmitted to them. 
However, we find that third-parties collect these unique identifiers more often 
than first-parties.
In fact, depending on the app permissions, third parties try to collect as many 
identifiers as they can: for instance, third-party domains like 
\emph{ad-x.co.uk}, 
\emph{adxtracking.com} and \emph{mobilecore.com} all collect and send the 
AndroidID, IMEI and WiFi-MAC address to their servers in clear-text.
Additionally, first and third-parties both send frequently these unique 
identifiers over the Internet to their servers unmodified 
(e.g., without hashing) and in clear-text (without SSL).
This is a serious threat to user privacy as a network eavesdropper can easily 
correlate the data flowing through the Internet. 
For interested raders, Table~\ref{table:sysIdentifiers_and_thirdParties_Android} 
in the appendix of this paper provides the whole list of servers and the 
corresponding unique identifiers transmitted to them by 140 Android apps 
in clear-text or through 
SSL.

Along with these unique system identifiers, third-parties also collect and send 
the names of apps in which their code is present. 
We notice that User-Agent http header field generally contains package/App name. 
As knowing the apps of a user reveals a lot about user 
interests~\cite{Seneviratne:2014} and increases the 
re-identification risk~\cite{achara2015}, this is a serious privacy threat.
In fact, the collection of such data is proportional to the number of apps in 
which third-party code is present and the number of apps sending these unique 
identifiers. 
So it is interesting to quantify the number 
of apps sending these unique identifiers to third-parties.
Looking at both Fig.~\ref{fig:Android_sysIdTransmitted} and 
Fig.~\ref{fig:Android_NoOfApps}, it can easily be 
deduced that the presence of third-parties in Android Apps is huge. 
Globally, we find that 31\% (44 out of 140) of Apps transmit, at least, 
one (un)modified unique identifier over the Internet.

\subsubsection{iOS.}

Fig.~\ref{fig:iOS_NoOfApps} presents the number of iOS apps (out of a total 
of 140 most popular apps tested) that transmit 
various unique system identifiers over the Internet. 
The ``AdIdentifier'' is transmitted the most, which is fine because it has been 
specifically introduced for Advertizing and Analytics purposes as a replacement 
to the deprecated UDID.
However we notice that some companies (e.g., tapjoyads.com, greystripe.com, 
mdotm.com, admob.com and ad-inside.com) are still using deprecated UDID.
We also found four apps collecting the device name. 
In iOS, the device name (DeviceName) is set by the user during the initial 
device setup and often contains the user's real name.
Since this device name is stable (the user generally does not modify it), 
even if it is not guaranteed to be unique across all devices, it is a stable 
identifer that can probably be used for tracking purposes.
Additionally, if it is set with the user's real name, it may reveal user 
identity.
We also notice that these identifiers are always collected when the user 
starts/stops interacting with the app. 
This means that third-parties can even know how long a user is using a 
particular app and the time when a user goes idle, revealing user 
habits. 
As, globally, 60\% (i.e., 84 out of 140) of apps send, at least, one 
(un)modified unique identifier over the Internet, the risk in terms of privacy 
is huge.

Fig.~\ref{fig:iOS_sysIdTransmitted} presents the number of servers where 
unique system identifiers are transmitted by 140 iOS apps.
We find a lot of apps transmitting AdIdentifier over the Internet to their 
servers as well as third-parties.
UDID is still being used and transmitted by a total of 9 apps. 
Surprisingly, the device name is transmitted to 5 servers. Out of these 5 
servers, only 1 server belongs to a first-party whereas others are either 
third-party or identified.
In case of iOS, Wi-Fi MAC address is the only hardware-tied identifier that is 
available to be accessed. It is transmitted to both first and third-parties.
More details about servers where these unique identifiers are sent, can be 
found in~Table~\ref{table:sysIdentifiers_and_thirdParties} in the appendix 
of this paper

\subsection{Unique identifiers generated by third-parties}

Let us now consider unique identifiers generated by third-parties in order to 
bypass OS restrictions.
As access to system unique identifier is limited on iOS, third-parties 
generate and share unique identifiers to have a mechanism to track users 
across apps.
However, apps are sandboxed on iOS (as well as on Android) and therefore, 
 these third-parties cannot simply share those generated unique identifiers across 
 apps. To circumvent this limitation on iOS, third-parties use a class called 
 UIPasteBoard~\cite{uipasteboard}. 
This is specifically designed for cut/copy/paste operations with the ability to 
share data between apps. 
The data shared by apps with this class can be persistently stored even across 
device restarts.  
Among the Apps we tested, we found that a large number of third-parties use the 
UIPasteboard class to share a unique third-party identifier across apps.  
Looking at the names and types of pasteboards created and the servers
where these values are sent, we found that 63 Apps create at least one
new pasteboard entry at the initiative of a third-party library
(Fig.~\ref{fig:iOS_NoOfApps}).

Essentially, third-party code present inside an application stores a pasteboard 
entry with its unique name, type and value. Later, if an App containing the code 
from the same third-party is installed, it retrieves the value corresponding to 
its pastebaord name. To have a look on pasteboard names, types and values used 
by various third-parties present inside 140 iOS Apps tested, please refer to 
Table~~\ref{table:listOfPBNamesTypesValues} in the appendix of this paper. Here
it is to be noted that user has no control over this kind of tracking and 
``Limit Ad Tracking'' feature of iOS is ineffective in this case.

From Fig.~\ref{fig:iOS_sysIdTransmitted}, we note that these pastboard 
entries are transmitted to various first and third-parties even though pastboard 
entries are just designed for cut/copy/paste operation between apps.
Moreover, we find that pastboard entries are more transmitted to 
third-parties than first-parties. This assures our assumption that third-parties 
generate these unique pastbaord entries to track users because they transmit 
them over the Internet.

\subsection{Comparation of third-party tracking on Android and iOS}

Comparing the identifiers transmitted by 140 most popular apps on Android 
and iOS reveals that iOS apps transmit system identifiers more often than Android 
apps.
However, iOS apps mostly transmit dedicated identifier, i.e., advertising id, 
for tracking and advertising purposes and not many apps transmit hardware-tied 
system identifiers as compared to app on Android.
This is probably due to the fact that many system identifiers (hardware and 
SIM-tied identifiers like IMEI, IMSI, and Serial number) are not available to 
be accessed on iOS.

As opposed to iOS apps, we did not notice Android apps generating and sharing 
unique systems identifiers for tracking purposes.
This is probably because various system identifiers are readily avaialble to be 
accessed by apps on Android and therefore, third-parites do not need to generate 
their own identifiers. 
Indeed, identifiers such as serial number and Android ID do not even need a permission to be accessed.
As tracking through system identifiers is more reliable 
and accurate, trackers probably do not need to resort to this solution.

In conclusion, Android makes available wider range of system unique
identifiers to apps as compared to iOS therefore it is easier for
third-parties to track Android users as compared to iOS. Nevertheless,
this does not stop trackers on iOS which resort to other techniques.

\section{Collection of other personal information}
\label{sec:PI_Collection}

To create a rich user profile, third-parties can use various means to collect a 
wide variety of personal information: 

\begin{enumerate}
\item By directly collecting as much information as possible from the device 
(i.e., by adding the appropriate code in the libraries to be included by the app 
developers).
\item By retrieving it from other third-parties who have already collected this 
information (thereby, aggregating the user PII~\cite{url:data_aggregation}).
\item By obtaining it from first-parties.
\end{enumerate}

It is difficult to measure how much information are being shared among 
third-parties themselves or among first and third-parties, but we next measure 
in this section what kinds of data and to which extent are being collected by 
these third-parties directly from the smartphone.

\subsection{Android}
Various personal data of the user is available to be accessed for apps on 
Android so that various useful apps can be developed.
Such personal data includes, e.g., location of the user, contacts or the accounts.
Our experiments with 140 Android apps reveal that different kind of user 
personal information is being sent over the Internet to various first and 
third-parties.

\begin{figure}[t!]  
  \centering
  \begin{subfigure}[]{0.75\linewidth}
    \scalebox{0.75}{\begin{tikzpicture}
\definecolor{col1}{RGB}{204, 204, 154}
\definecolor{col2}{RGB}{136, 136, 136}
\pgfplotsset{compat=newest}
\pgfplotsset{major grid style={gray!50}}
\begin{axis}[
ybar,
 every axis plot post/.style={/pgf/number format/fixed},
            bar width=8pt,
y tick label style={major tick length=0pt},
ymajorgrids,
axis lines*=none,
 axis line style={ultra thin,white},
            xtick=data,
enlargelimits=0.15,
ylabel={\#Apps},
symbolic x coords={NetworkCode,OperatorName,Location,Accounts,WiFiAPInfo,Contacts},
xtick=data,
nodes near coords,
nodes near coords align={vertical},
x tick label style={rotate=45,anchor=east},
ymin=0,ymax=20,
]
\addplot[draw=none,fill=col1, draw opacity=0,drop shadow={shadow yshift=0.4pt, shadow xshift=0.4pt, opacity=0.50}] coordinates {(NetworkCode,17)(OperatorName,16)(Location,6)(Accounts,3)(WiFiAPInfo,2)(Contacts,1)};
\end{axis}
\end{tikzpicture}}
    \caption{Android. \label{fig:Android_NoOfApps_PI}}
  \end{subfigure}
  \begin{subfigure}[]{0.75\linewidth}
    \scalebox{0.75}{\begin{tikzpicture}
\definecolor{col1}{RGB}{204, 204, 154}
\definecolor{col2}{RGB}{136, 136, 136}
\pgfplotsset{compat=newest}
\pgfplotsset{major grid style={gray!50}}
\begin{axis}[
ybar,
every axis plot post/.style={/pgf/number format/fixed},
            bar width=8pt,
y tick label style={major tick length=0pt},
ymajorgrids,
axis lines*=none,
 axis line style={ultra thin,white},
            xtick=data,
enlargelimits=0.15,
legend style={at={(0.7,0.95)},
anchor=north,legend columns=-1},
ylabel={\#Apps},
symbolic x coords={OperatorName,Location,DeviceName,Accounts,Contacts,SIM No},
xtick=data,
nodes near coords,
nodes near coords align={vertical},
x tick label style={rotate=45,anchor=east},
ymin=0,ymax=20,
]
\addplot[draw=none,fill=col1, draw opacity=0,drop shadow={shadow yshift=0.4pt, shadow xshift=0.4pt, opacity=0.50}] coordinates {(OperatorName,13)(Location,10)(DeviceName,4)(Accounts,2)(Contacts,1)(SIM No,1)};
\end{axis}
\end{tikzpicture}}
    \caption{iOS.\label{fig:iOS_NoOfApps_PI}}
  \end{subfigure}
  \caption{\# Apps sending PII out of a total of 140 apps.  \label{fig:iOS_Android_NoOfApps_PI}}
\end{figure}
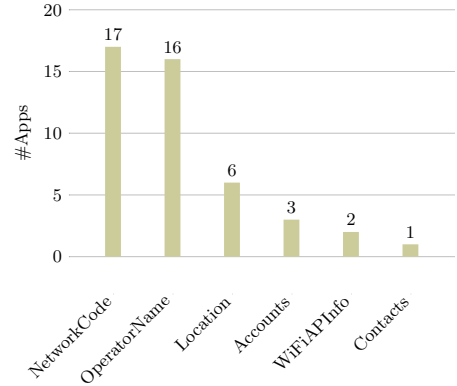
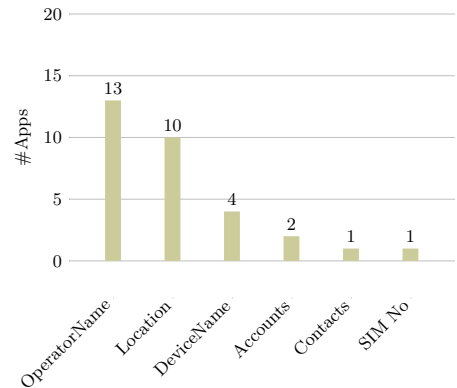

Fig.~\ref{fig:Android_NoOfApps_PI} presents number of apps sending different 
kind of data over the Internet. We see that network code and operator name is 
sent by 17 and 16 apps respectively. User location and accounts information 
is transmitted by six and three apps respectively. Two apps transmit 
information related to the Wi-Fi access point they are connected to whereas one 
app transmit user contacts over the Internet.

\begin{figure}[!hbt]  
\centering
	\begin{subfigure}[]{0.75\linewidth}
		\scalebox{0.75}{\begin{tikzpicture}
\definecolor{Col1}{HTML}{CC0000}
\definecolor{Col2}{HTML}{CCCC99}
\definecolor{Col3}{HTML}{003366}
\definecolor{Col4}{HTML}{996600}
\definecolor{Col6}{HTML}{669966}
\definecolor{Col7}{HTML}{666699}
\definecolor{Col8}{HTML}{FFCC00}
\begin{axis}[
ybar stacked,
ymajorgrids,
ymin=0,
axis line style={draw=none},
y tick label style={major tick length=0pt},
x tick label style={major tick length=0pt},
myplotShadow/.style={drop shadow={shadow yshift=0.25pt, shadow xshift=0.65pt}},
myplot/.style={draw=none,area legend,draw opacity=0,myplotShadow},
enlargelimits=0.05,
legend style={at={(0.7,0.95)},
anchor=north,legend columns=-1},
ylabel={\# Servers},
symbolic x coords={Accounts,Contacts,Location,Operator name,SIM network code,Wi-Fi scan/config},
xtick=data,
nodes near coords,
nodes near coords align={vertical},
x tick label style={rotate=45,anchor=east},
ymin=0,ymax=20,
]
\addplot[myplot,fill=Col3] coordinates {(Accounts,2)(Contacts,1)(Location,3)(Operator name,5)(SIM network code,3)(Wi-Fi scan/config,1)};
\addplot[myplot,fill=Col2] coordinates {(Accounts,1)(Contacts,0)(Location,9)(Operator name,8)(SIM network code,9)(Wi-Fi scan/config,1)};
\addplot[myplot,fill=Col4] coordinates {(Accounts,1)(Contacts,0)(Location,0)(Operator name,2)(SIM network code,4)(Wi-Fi scan/config,0)};
\end{axis}
\end{tikzpicture}}
		\caption{Android.\label{fig:Android_piiTransmitted}}
	\end{subfigure}
	\begin{subfigure}[]{0.75\linewidth}
		\scalebox{0.75}{\begin{tikzpicture}
\definecolor{Col1}{HTML}{CC0000}
\definecolor{Col2}{HTML}{CCCC99}
\definecolor{Col3}{HTML}{003366}
\definecolor{Col4}{HTML}{996600}
\definecolor{Col6}{HTML}{669966}
\definecolor{Col7}{HTML}{666699}
\definecolor{Col8}{HTML}{FFCC00}
\begin{axis}[
ybar stacked,
ymajorgrids,
ymin=0,
axis line style={draw=none},
y tick label style={major tick length=0pt},
x tick label style={major tick length=0pt},
myplotShadow/.style={drop shadow={shadow yshift=0.25pt, shadow xshift=0.65pt}},
myplot/.style={draw=none,area legend,draw opacity=0,myplotShadow},
enlargelimits=0.05,
legend style={at={(0.7,0.95)},
anchor=north,legend columns=-1},
ylabel={\# Servers},
symbolic x coords={Accounts,Contacts,Location,SIM network name,SIM number,DeviceName},
xtick=data,
nodes near coords,
nodes near coords align={vertical},
x tick label style={rotate=45,anchor=east},
ymin=0,ymax=20,
]
\addplot[myplot,fill=Col3] coordinates {(Accounts,1)(Contacts,1)(Location,3)(SIM network name,2)(SIM number,1)(DeviceName,1)};
\addplot[myplot,fill=Col2] coordinates {(Accounts,0)(Contacts,0)(Location,2)(SIM network name,4)(SIM number,1)(DeviceName,2)};
\addplot[myplot,fill=Col4] coordinates {(Accounts,0)(Contacts,0)(Location,0)(SIM network name,0)(SIM number,0)(DeviceName,0)};
\end{axis}
\end{tikzpicture}}
		\caption{iOS.\label{fig:iOS_piiTransmitted}}
	\end{subfigure}\\
	\scalebox{0.75}{\definecolor{Col1}{HTML}{CC0000}
\definecolor{Col2}{HTML}{CCCC99}
\definecolor{Col3}{HTML}{003366}
\definecolor{Col4}{HTML}{996600}
\definecolor{Col6}{HTML}{669966}
\definecolor{Col7}{HTML}{666699}
\definecolor{Col8}{HTML}{FFCC00}
\begin{tikzpicture}
    \begin{customlegend}[legend columns=5,legend style={nodes={right},draw=none,font=\small},legend entries={First-party, Third-party, Unidentified}]
        \addlegendimage{draw=none,area legend,draw opacity=0,blur shadow={shadow blur radius=0.25pt,shadow yshift=-0.5pt, shadow xshift=0.5pt},fill=Col3}
        \addlegendimage{draw=none,area legend,draw opacity=0,blur shadow={shadow blur radius=0.25pt,shadow yshift=-0.5pt, shadow xshift=0.5pt},fill=Col2}
        \addlegendimage{draw=none,area legend,draw opacity=0,blur shadow={shadow blur radius=0.25pt,shadow yshift=-0.5pt, shadow xshift=0.5pt},fill=Col4}
        \addlegendimage{only marks, mark=square*,draw opacity=0,fill=predictedCol,legend image post style={xshift=0.2cm}}
    \end{customlegend}
\end{tikzpicture}}
	\caption{\# servers where user personal information is sent by 140 apps.\label{fig:iOS_Android_pii_transmitted}}
\end{figure}
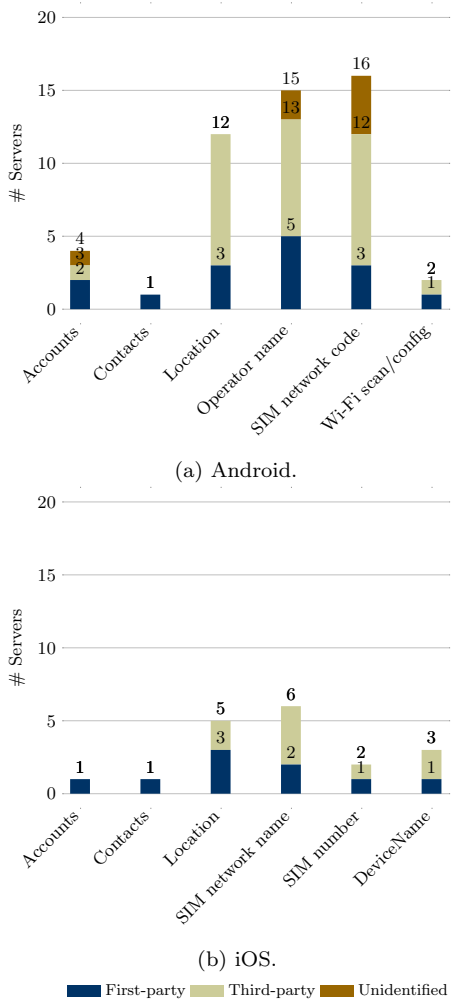

Fig.~\ref{fig:Android_piiTransmitted} reveal that user location is transmitted 
(encrypted or in clear-text) to nine third-parties whereas to three first parties.
In fact, it is more often sent to third-parties than first-parties. 
We also find that user location is sent (encrypted or in clear-text) to nine 
third-parties whereas it is sent to only three first-parties. 
This means that user location is used more often for tracking and profiling the 
user and not for providing a useful service.
Otherwise, we also note that the name of the telephony operator and the SIM 
network code is being collected by a lot of first and third-parties 
(details available in Table~\ref{table:OtherPrivDataSentToWhere_Android} in appendix).

Moreover, as the apps installed on a device is highly valuable information for 
trackers/advertisers to infer user interests and habits, we detect and measure 
the leakage of this information too. 
We find that 5 third-parties know 4 or more apps installed by a user. 
This puts users at serious privacy risks as users can be re-identified later 
with high probablity~\cite{achara2015}.
Specifically, ``tardemob.com", present in ``Booking.com" App, collects the list 
of all apps installed on the device and sends this list to its server
(details available in Table~\ref{table:ThirdPartyWithProcessNamesKnown_Android} in appendix).

\subsection{iOS} 
iOS also makes accessible many user PII sources (e.g., Accounts, Location, or Contacts) to apps. 
This is necessary so that a wide variety of apps can be developed.

Figure~\ref{fig:iOS_NoOfApps_PI} presents the number of apps sending user PII
over the Internet. We find that 10 apps (out of 140) send user location over 
the Internet.  SIM network (operator) name and SIM number are transmitted most 
by iOS apps.
Device name is transmitted by four apps.

Fig.~\ref{fig:iOS_piiTransmitted} shows whether the location data 
it is sent to first or third-parties by these 10 apps.
We find that it is sent to two third-parties (to one in clear-text and one 
using SSL) and three first-parties. SIM network name is mostly transmitted 
to third-parties
(details about the transmitted user data and where it is sent to are
available in Table~\ref{table:OtherPrivDataSentToWhere_iOS} in the
appendix).

\subsection{Comparision between Android and iOS}

We find that 10 apps (out of 140) send user location over the Internet
as compared to 6 apps on Android.  However, more third-parties collect
and send user location over the Internet on Android. This means that
on iOS, there are less third-parties but they are more broadly used by
apps, on the other hand there are more third-parties in Android used
by less apps.

As opposed to Android, we find that iOS apps leak less user PII. 
In total, there are 8 (as opposed to 21 on Android) third-parties where user PII 
is sent to on iOS.
Also, both first and third-parties did not send much data to their servers in 
clear-text. There is only one third-party server and one first-party server where 
user location is sent in clear-text on iOS as compared to six and one respectively 
on Android. 
Globally, we note that iOS apps sent lesser user PII over the Internet as they use
SSL more often than their Android counterparts.

iOS apps also leak more information about the list of 
installed apps on the phone as compared to Android apps. 
Nine third-parties know, at least, five names of the installed packages. 
Flurry, for example, knows 25 apps installed on the phone and is included in all 
these apps. The included library from Flurry sends the name of the app in which it 
is present as part of the communication with their servers.
Moreover, the collection of this information is in plain-text.
To get the complete list of these third-parties as well as the package names 
known to them, please refer to Table~\ref{table:ThirdPartyWithProcessNamesKnown} 
in the appendix of the paper.

\section{Discussion}
\label{sec:Efficacy_of_Safeguards}

\subsection{Comparing MobileAppScrutinator on Android with TaintDroid}

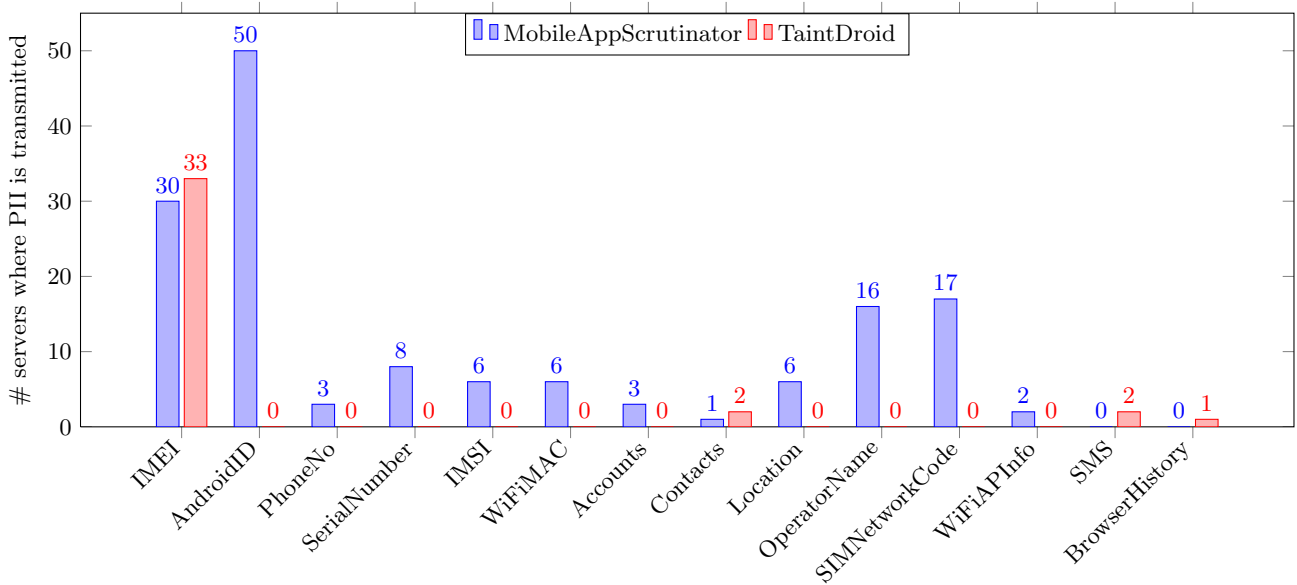
\begin{figure*}[!t]
\centering
\pgfplotstableread[row sep=\\,col sep=&]{
    pii & appscrutinator & taintdroid \\
    IMEI     & 30  & 33    \\
    AndroidID     & 50 & 0    \\
    PhoneNo    & 3 & 0  \\
    SerialNumber  & 8 & 0  \\
    IMSI   & 6  & 0  \\
    WiFiMAC      & 6  & 0  \\
    Accounts     & 3  & 0    \\
    Contacts     & 1 & 2    \\
    Location    & 6 & 0  \\
    OperatorName  & 16 & 0  \\
    SIMNetworkCode   & 17  & 0  \\
    WiFiAPInfo      & 2  & 0  \\
    SMS   & 0  & 2  \\
    BrowserHistory      & 0  & 1  \\
    }\mydata

\begin{tikzpicture}
    \begin{axis}[
            ybar,
            bar width=.3cm,
            width=\textwidth,
            height=.4\textwidth,
            legend style={at={(0.5,1)},
                anchor=north,legend columns=-1},
            symbolic x coords={IMEI,AndroidID,PhoneNo,SerialNumber,IMSI,WiFiMAC,Accounts,Contacts,Location,OperatorName,SIMNetworkCode,WiFiAPInfo,SMS,BrowserHistory},
            xtick=data,
            nodes near coords,
            nodes near coords align={vertical},
            ymin=0,ymax=55,
            ylabel={\# servers where PII is transmitted},
            x tick label style={rotate=45,anchor=east},
        ]
        \addplot table[x=pii,y=appscrutinator]{\mydata};
        \addplot table[x=pii,y=taintdroid]{\mydata};
        \legend{MobileAppScrutinator, TaintDroid}
    \end{axis}
\end{tikzpicture}
\caption{Comparation of privacy leaks reported by MobileAppScrutinator on 
Android and TaintDroid}
\label{fig:taintdroidVSappscrutinator}
\end{figure*}

Fig.~\ref{fig:taintdroidVSappscrutinator} compares privacy leaks reported by 
MobileAppScrutinator on Android and TaintDroid for some personal data.
To know more details, please refer to the Table~\ref{table:pii_taintdroid} in 
the appendix of this paper that presents the PII 
(unique identifiers and other private information) leaks reported by running the 
same set of applications on TaintDroid 4.3.
We note that TaintDroid only reported the leakage of one unique system identifier 
(IMEI) whereas MobileAppScrutinator reports the leakge of six different unique 
identifiers.
Moreover, MobileAppScrutinator overall reported more privacy leaks than
TaintDroid.
However, we interestingly found that both MobileAppScrutinator and TaintDroid 
have false negatives.
This suggests that these tools should not be replacement of one another but can 
actually be complimentary to each other.

While comparing PII leaks reported by MobileAppScrutinator with TaintDroid, we 
found that TaintDroid did not report any leakage of location coordinates. 
TaintDroid reported leakge of Address Book (Contacts) information to two 
third-parties whereas MobileAppScrutinator could only detect Contacts leakge to 
only one party. Again it is intersting to note here that parties, where Contacts
 information was leaked, are mutually exclusive. TaintDroid also reported the 
 leakage of Browser History and SMS to one and two parties respectively but, as  
 MobileAppScrutinator did not implement their leakge detection, we cannot 
 compare with respect to these two types of PII.
On the contrary, MobileAppScrutinator was implemented to detect the leakage of other 
 kinds of private data, such as Accounts, Operator Name, SIM Network Code and 
 WiFi Scan/Config info, which current implementaion of TaintDroid (version 4.3) 
 lacked.

\subsection{Effectiveness of varous privacy safeguards}

In order to provide transparency and control over privacy, both Android and iOS 
involve user decisions along with mechanisms adopted by their respective systems. 
However, the approach followed by Android and iOS is different: Android employs 
a static install-time permission system whereas iOS solicits explicit user 
permission at runtime. 
No doubt these OS mechanisms are mostly effective, they lack behavirol analysis, 
i.e., when, where and how often the accessed information is sent over network. 
For example, it is vital to distinguish the fact if the PII is sent to an 
application server or to a remote third-party.
In fact, a user giving access to her PII for a desired service does not 
necessarily mean that she also wants to share this information with other 
parties, for example, advertisers or analytics companies. 
Similarly, an application accessing and sending user location only at 
installation time is not the same as sending it every five minutes. 

Below we discuss the effectiveness of various privacy safeguards avaialabe on 
both Android and iOS based on our experiments and results.

\paragraph{Resetting the ``AdIdentifier" on iOS} 
The effect of resetting the AdIdentifier is not similar to 
``Deleting the cookies" in web tracking and could easily be nullified.
Resetting the AdIdentifier, in theory, is meant to prevent trackers from
linking the user activity before and after the reset.  However the trackers can
easily detect the AdIdentifier change and link the two values even if Apple 
explicitly tells not to do so. Apps are not technically restricted by iOS to 
respect the resetting of AdIdentifier on iOS. In our study, we find that 20\% 
apps send the IdentifierForVendor\footnote{This is a stable identifier unique to 
all apps of a single developer on a particular device, that cannot be changed or 
reset by the user.} along with the AdIdentifier to third-parties 
(Details in Table~\ref{table:ServersOfIdForVendor} in appendix). 
It is noticeable that many third-parties collect
this identifier, whereas it was principally designed by Apple to be used only
by first-parties. As IdentifierForVendor is being collected by third-parties, 
they are able to link the AdIdentifiers before and after the reset.

\paragraph{Apps bypass the ``AdIdentifier" on iOS}
We have seen that many apps are using other tracking mechanisms to track the
user in addition to the AdIdentifier. In our experiments, we discovered that 93
apps out of 140 (i.e., approx. 66\%) will continue to track the user after a
reset of the AdIdentifier by the user. This measurement does not even 
consider the applications employing the previously described technique to match
the changed/reset AdIdentifiers as we cannot be sure what third-parties do with 
their data collected.  In iOS 7, Apple banned the access to WiFi MAC Address, 
but the percentage only reduces from 66\% to approximately 42\%
(60 Apps out of a total of 140 Apps.), i.e., if we exclude the apps (24\%) using 
only WiFi MAC address as a unique identifier for tracking.

 \paragraph{``Limit Ad Tracking'' iOS privacy setting} 
 It turns out that the ``AdIdentifier'' is available to all Apps, even after a
 user has chosen the ``Limit Ad Tracking'' option. It is therefore ambiguous how
 iOS can enforce the ``Limit Ad Tracking'' option. In fact, ``Limit Ad Tracking'' 
 setting appear to be misleading, as it gives the end-user a wrong feeling of 
 opting-out from device tracking.  And if
 recently Apple started to reject apps accessing the AdIdentifier without
 providing In-App Advertising~\cite{url:apple_adId_based_rejection}, which is
 reasonable, it still does not solve the core problem.

\section{Conclusion}

This paper first introduces the MobileAppScrutinator platform for the study of
third-party smartphone tracking.  To the best of our knowledge, this platform
is the first one that embraces both iOS and Android, using the same
dynamic analysis methodology in both cases. For the first time, it provides
in-depth insight on the different PII accessed, hashed and/or encrypted
and sent to remote servers, either in clear-text or over SSL connections.  
This in-depth analysis capability is a key to analyze the applications
and understand how they (ab)use personal information.

The second major contribution of this work is the behavioral analysis, thanks
to the MobileAppScrutinator platform, of 140 free and popular apps, selected so that
they are available on both mobile OSs in order to enable comparisons. Two
important aspects are considered: first we show that many stable identifiers
are collected on Android, in order to track individual
devices in the long term. 
On iOS, availability of system-level identifiers is less common, but techniques have been designed to create new cross-app,
stable identifiers by third-parties themselves.  The second aspect concerns the
user-related information. We show that a significant amount of PII is being
collected by third-parties who implicitly know a lot about the user interests
(e.g., by collecting the list of apps installed or currently running).

Finally, this work enables to have a comparative view of ongoing
tracking on Andorid and iOS. Our experiments show that Android apps
are more privacy-invasive when compared to iOS apps as the presence of
third-parties is clearly more frequent in Android applications. In all
cases, protective measures should be taken by device manufacturers, OS
designers and various regulatory authorities in a coordinated way to
control the collection and usage of PII.

\bibliographystyle{IEEEtran}
\bibliography{paper}

\appendix
\begin{table*}
\small
\centering
\caption{Unique System Identifiers transmitted by 140 Android Apps tested.}
\label{table:sysIdentifiers_and_thirdParties_Android}
\resizebox{2\columnwidth}{!}{
\begin{tabular}{|c|c||c|c|c|c|c|c|c|c|c|}
\cline{3-11}
\multicolumn{2}{c|}{}&\multirow{2}{*}{\textbf{Server}}&\multicolumn{2}{c|}{\textbf{AndroidID}}&\multirow{2}{*}{\textbf{PhoneNo}}&\multicolumn{2}{c|}{\textbf{IMEI}}&\multirow{2}{*}{\textbf{SerialNo}}&\multirow{2}{*}{\textbf{IMSI}}&\multirow{2}{*}{\textbf{WiFi MAC}} \\ \cline{4-5} \cline{7-8} 
\multicolumn{2}{c|}{}&&\textbf{Modified}&\textbf{Unmodified}&&\textbf{Modified}&\textbf{Unmodified}&&& \\ \hline \hline

\multirow{29}{*}{\rotatebox{90}{\textbf{Third-parties}}}&\multirow{18}{*}{\rotatebox{90}{\textbf{Clear}}}&amazonaws.com&\checkmark&\checkmark&&\checkmark&&&&\checkmark \\ \cline{3-11}

&&ad-x.co.uk&&\checkmark&&&\checkmark&&&\checkmark \\ \cline{3-11}
&&mobilecore.com&&\checkmark&&&\checkmark&&&\checkmark \\ \cline{3-11}

&&kochava.com&&\checkmark&&&\checkmark&& & \\ \cline{3-11}
&&apsalar.com&\checkmark&\checkmark&&&&&& \\ \cline{3-11}
&&mdotm.com&&\checkmark&&&\checkmark&&& \\ \cline{3-11} 
&&adtilt.com&&\checkmark&&&\checkmark&&& \\ \cline{3-11}
&&estat.com&&&&&\checkmark&\checkmark&&\\ \cline{3-11}

&&sophiacom.fr&&\checkmark&&&&&& \\ \cline{3-11}
&&appnext.com&&\checkmark&&&&&& \\ \cline{3-11}
&&flurry.com&&\checkmark&&&&&& \\ \cline{3-11}
&&socialquantum.ru&&&&&&&&\checkmark \\ \cline{3-11}
&&sitestat.com&&&&&&\checkmark&& \\ \cline{3-11}
&&pureagency.com&&&&&\checkmark&&& \\ \cline{3-11}
&&smartadserver.com&&\checkmark&&&&&& \\ \cline{3-11}
&&xiti.com&&\checkmark&&&&&& \\ \cline{3-11}
&&playhaven.com&&\checkmark&&&&&& \\ \cline{3-11}
&&yoz.io&&\checkmark&&&&&& \\ \cline{3-11}
&&seattleclouds.com&&\checkmark&&&&&& \\ \cline{3-11}
&&ad-market.mobi&&\checkmark&&&&&& \\ \cline{2-11}

&\multirow{11}{*}{\rotatebox{90}{\textbf{SSL}}}&tapjoyads.com&&\checkmark&&\checkmark&\checkmark&\checkmark&&\checkmark \\ \cline{3-11}

&&airpush.com&\checkmark&\checkmark&\checkmark&\checkmark&&&& \\ \cline{3-11}

&&revmob.com&&\checkmark&&&\checkmark&\checkmark&& \\ \cline{3-11}
&&appwiz.com&&\checkmark&\checkmark&&\checkmark&&& \\ \cline{3-11}

&&amazon.com&&\checkmark&&&&\checkmark&& \\ \cline{3-11}
&&adcolony.com&\checkmark&&&\checkmark&&&& \\ \cline{3-11}
&&fiksu.com&&\checkmark&&&\checkmark&&& \\ \cline{3-11}
&&crittercism.com&&\checkmark&&&\checkmark&&& \\ \cline{3-11}
&&googleapis.com&&&\checkmark&&&&\checkmark& \\ \cline{3-11}
&&appsflyer.com&&&&&\checkmark&&&\checkmark \\ \cline{3-11}

&&dataviz.com&&&&&\checkmark&&& \\ \cline{3-11}
&&mobileapptracking.com&&\checkmark&&&&&& \\ \cline{1-11}

\multirow{13}{*}{\rotatebox{90}{\textbf{First-parties}}}&\multirow{5}{*}{\rotatebox{90}{\textbf{Clear}}}&mobage.com&&\checkmark&&&\checkmark&&& \\ \cline{3-11}

&&ijinshan.com&&&&&\checkmark&&& \\ \cline{3-11}
&&blitzer.de&&&&&\checkmark&&& \\ \cline{3-11}
&&eurosport.com&&&&&&&\checkmark& \\ \cline{3-11}
&&cdiscount.com&&&&&&&\checkmark& \\ \cline{2-11}

&\multirow{8}{*}{\rotatebox{90}{\textbf{SSL}}}&google.com&&\checkmark&&&\checkmark&\checkmark&& \\ \cline{3-11}
&&badoo.com&&\checkmark&&&\checkmark&&\checkmark& \\ \cline{3-11}

&&dropbox.com&&\checkmark&&&&\checkmark&& \\ \cline{3-11}

&&klm.com&&\checkmark&&&&&& \\ \cline{3-11}
&&airfrance.com&&\checkmark&&&&&& \\ \cline{3-11}
&&airbnb.com&&\checkmark&&&&&& \\ \cline{3-11}
&&groupon.com&&\checkmark&&&&&& \\ \cline{3-11}
&&adobe.com&&&&&&&\checkmark& \\ \cline{1-11}

\multirow{18}{*}{\rotatebox{90}{\textbf{Unidentified}}}&\multirow{16}{*}{\rotatebox{90}{\textbf{Clear}}}&72.21.194.112&&\checkmark&&&&\checkmark&& \\ \cline{3-11}
&&dxsvr.com&&&&&\checkmark&&\checkmark& \\ \cline{3-11}
&&69.28.52.39&&\checkmark&&&\checkmark&&& \\ \cline{3-11}
&&198.61.246.5&&\checkmark&&&\checkmark&&& \\ \cline{3-11}
&&183.61.112.40&&\checkmark&&&\checkmark&&& \\ \cline{3-11}
&&61.145.124.113&&\checkmark&&&\checkmark&&& \\ \cline{3-11}

&&74.217.75.7&&\checkmark&&&&&& \\ \cline{3-11}
&&183.61.112.40&&&&&&&\checkmark& \\ \cline{3-11}
&&linode.com&&&&\checkmark&&&& \\ \cline{3-11}
&&93.184.219.20&\checkmark&&&&&&& \\ \cline{3-11}
&&107.6.111.137&\checkmark&&&&&&& \\ \cline{3-11}
&&startappexchange.com&&\checkmark&&&&&& \\ \cline{3-11}
&&91.103.140.6&&\checkmark&&&&&& \\ \cline{3-11}
&&209.177.95.171&&\checkmark&&&&&& \\ \cline{3-11}
&&ati-host.net&\checkmark&&&&&&& \\ \cline{3-11}
&&adkmob.com&&\checkmark&&&&&& \\ \cline{2-11}

&\multirow{2}{*}{\rotatebox{90}{\textbf{SSL}}}&fastly.net&&\checkmark&&&&&& \\ \cline{3-11}
&&canal-off.sbw-paris.com&&&&&\checkmark&&& \\ \cline{1-11}
\end{tabular}
}
\end{table*}

\begin{table}[h]
\caption{Detection of PII leakage using TaintDroid 4.3 (Same applications are tested to compare the results with MobileAppScrutinator platform).}
\label{table:pii_taintdroid}
\resizebox{\columnwidth}{!}{
\begin{tabular}{|c|c||c|c|c|c|c|}
\cline{3-7}
\multicolumn{2}{c|}{}&\textbf{Server}&\textbf{IMEI}&\parbox[c]{1.5cm}{\textbf{Browser History}}&\parbox[c]{1.5cm}{\textbf{Address Book}}&\textbf{SMS}\\ \hline \hline

\multirow{15}{*}{\rotatebox{90}{\textbf{Third-parties}}}&\multirow{11}{*}{\rotatebox{90}{\textbf{Clear}}}&hinet.net&\checkmark&&&\\ \cline{3-7} 
&&enovance.net&\checkmark&&& \\ \cline{3-7}
&&aol.com&\checkmark&&& \\ \cline{3-7}
&&kimsufi.com&\checkmark&&& \\ \cline{3-7}
&&typhone.net&\checkmark&&& \\ \cline{3-7}
&&betacie.net&\checkmark&&& \\ \cline{3-7}
&&1e100.net&\checkmark&&& \\ \cline{3-7}
&&dolphin-server.co.uk&\checkmark&&& \\ \cline{3-7}
&&linode.com&\checkmark&&& \\ \cline{3-7}
&&amazon.com&\checkmark&&& \\ \cline{3-7}
&&ati-host.net&\checkmark&&& \\ \cline{2-7}
&\multirow{4}{*}{\rotatebox{90}{\textbf{SSL}}}&amazonaws.com&\checkmark&&\checkmark& \\ \cline{3-7}

&&1e100.net&&&\checkmark& \\ \cline{3-7}
&&skyhookwireless.com&\checkmark&&& \\ \cline{3-7}
&&akamaitechnologies.com&\checkmark&&& \\ \cline{1-7}

\multirow{5}{*}{\rotatebox{90}{\textbf{First-parties}}}&\multirow{3}{*}{\rotatebox{90}{\textbf{Clear}}}&teamviewer.com&\checkmark&&& \\ \cline{3-7}

&&badoo.com&\checkmark&&& \\ \cline{3-7}
&&shazamteam.net&\checkmark&&& \\ \cline{2-7}
&\multirow{2}{*}{\rotatebox{90}{\textbf{SSL}}}&svcs.paypal.com&\checkmark&&& \\ \cline{3-7}

&&amazon.com&\checkmark&&& \\ \cline{1-7}

\multirow{17}{*}{\rotatebox{90}{\textbf{Unidentified}}}&\multirow{9}{*}{\rotatebox{90}{\textbf{Clear}}}&162.13.174.5&\checkmark&&& \\ \cline{3-7}
&&69.28.52.38&\checkmark&&& \\ \cline{3-7}
&&195.154.141.2&\checkmark&&& \\ \cline{3-7}
&&188.165.90.225&\checkmark&&& \\ \cline{3-7}
&&91.103.140.225&\checkmark&&& \\ \cline{3-7}
&&61.145.124.113&\checkmark&&& \\ \cline{3-7}
&&69.28.52.36&&&&\checkmark \\ \cline{3-7}
&&183.61.112.40&&&&\checkmark \\ \cline{3-7}
&&91.213.146.11&\checkmark&&& \\ \cline{2-7}
&\multirow{8}{*}{\rotatebox{90}{\textbf{SSL}}}&31.222.69.213&\checkmark&&& \\ \cline{3-7}
&&212.31.79.7&\checkmark&&& \\ \cline{3-7}
&&92.52.84.202&\checkmark&&& \\ \cline{3-7}
&&69.194.39.80&\checkmark&&& \\ \cline{3-7}
&&72.26.211.237&\checkmark&&& \\ \cline{3-7}
&&192.225.158.1&\checkmark&&& \\ \cline{3-7}
&&54.256.81.235&\checkmark&&& \\ \cline{3-7}
&&67.222.111.117&&\checkmark&& \\ \cline{1-7}
\end{tabular}
}
\end{table}

\begin{table*}
\small
\centering
\caption{Unique System Identifiers transmitted by 140 iOS Apps tested.}
\label{table:sysIdentifiers_and_thirdParties}
\resizebox{2\columnwidth}{!}{
\begin{tabular}{|c|c|c|c|c|c|c|c|c|}
\cline{3-9}
\multicolumn{2}{c|}{}&\multirow{2}{*}{\textbf{Server}}&\multirow{2}{*}{\textbf{AdIdentifier}}&\multirow{2}{*}{\textbf{UDID}}&\multirow{2}{*}{\textbf{DeviceName}}&\multicolumn{2}{c|}{\textbf{WiFi MAC}}&\multirow{2}{*}{\textbf{Pasteboard IDs}}\\ \cline{7-8} 
\multicolumn{2}{c|}{}&&&&&\textbf{Modified}&\textbf{Unmodified}& \\ \hline \hline
\multirow{31}{*}{\rotatebox{90}{\textbf{Third-parties}}}&\multirow{15}{*}{\rotatebox{90}{\textbf{Clear}}}&clara.net&\checkmark&&&&\checkmark&\checkmark \\ \cline{3-9}

&&appads.com&\checkmark&&&&& \\ \cline{3-9}
&&amazonaws.com&&&&&&\checkmark \\ \cline{3-9}
&&1e100.net&&&&&&\checkmark \\ \cline{3-9}
&&adcolony.com&\checkmark&&&&& \\ \cline{3-9}
&&facebook.com&&&&&&\checkmark \\ \cline{3-9}
&&your-server.de&&&&&&\checkmark \\ \cline{3-9}
&&sophiacom.fr&\checkmark&&&&& \\ \cline{3-9}
&&smartadserver.com&\checkmark&&&&& \\ \cline{3-9}
&&mopub.com&\checkmark&&&&& \\ \cline{3-9}
&&sofialys.net&&&&&&\checkmark \\ \cline{3-9}
&&visuamobile.com&&&&&&\checkmark \\ \cline{3-9}
&&swelen.com&\checkmark&&&&& \\ \cline{3-9}
&&adtilt.com&\checkmark&&&&& \\ \cline{3-9}
&&nanigans.com&\checkmark&&&&& \\ \cline{3-9}
&&tapjoyads.com&&\checkmark&&&& \\ \cline{3-9} 
&&greystripe.com&&\checkmark&&&& \\ \cline{3-9}
&&mdotm.com&&\checkmark&&&& \\ \cline{3-9}
&&sofialys.net&&&&&&\checkmark \\ \cline{3-9}
&&visuamobile.com&&&&&&\checkmark \\ \cline{3-9}
&&admob.com&&\checkmark&&&& \\ \cline{3-9}
&&ad-inside.com&&\checkmark&&&& \\ \cline{3-9}
&&xiti.com&&&&&&\checkmark \\ \cline{3-9}
&&2o7.net&&&&&&\checkmark \\ \cline{3-9}
&&jumptap.com&\checkmark&&&&& \\ \cline{2-9}

&\multirow{16}{*}{\rotatebox{90}{\textbf{SSL}}}&tapjoyads.com&\checkmark&&&&\checkmark& \\ \cline{3-9}
&&trademob.net&\checkmark&&&&\checkmark& \\ \cline{3-9}
&&adjust.io&\checkmark&&&&& \\ \cline{3-9}
&&boxcar.io&&&\checkmark&&& \\ \cline{3-9}
&&flurry.com&&&&\checkmark&& \\ \cline{3-9}
&&tapjoy.com&\checkmark&&&&& \\ \cline{3-9}
&&mobile-adbox.com&\checkmark&&&&& \\ \cline{3-9}
&&fiksu.com&\checkmark&&&&&\checkmark \\ \cline{3-9}
&&tapad.com&\checkmark&&&&& \\ \cline{3-9}
&&testflightapp.com&\checkmark&&&&& \\ \cline{3-9}
&&adtilt.com&\checkmark&&&&& \\ \cline{3-9}
&&nanigans.com&\checkmark&&&&& \\ \cline{3-9}
&&ad-x.co.uk&\checkmark&&&&& \\ \cline{3-9}
&&crittercism.com&&&\checkmark&&& \\ \cline{3-9}
&&facebook.com&\checkmark&&&&& \\ \cline{3-9}
&&newrelic.com&&&&&&\checkmark \\ \cline{3-9}
&&adzcore.com&\checkmark&&&&& \\ \cline{1-9}

\multirow{7}{*}{\rotatebox{90}{\textbf{First-parties}}}&\multirow{2}{*}{\rotatebox{90}{\textbf{Clear}}}&gameloft.com&\checkmark&&&&\checkmark&\checkmark \\ \cline{3-9}
&&spotify.com&&&&&&\checkmark \\ \cline{3-9}
&&disneyis.com&&&&&&\checkmark \\ \cline{3-9}
&&mobiata.com&&\checkmark&&&& \\ \cline{2-9}

&\multirow{5}{*}{\rotatebox{90}{\textbf{SSL}}}&gameloft.com&\checkmark&&&&\checkmark& \\ \cline{3-9}
&&paypal.com&\checkmark&&\checkmark&&&\checkmark \\ \cline{3-9}
&&booking.com&\checkmark&&&&& \\ \cline{3-9}
&&eamobile.com&\checkmark&&&&& \\ \cline{3-9}
&&expedia.com&&\checkmark&&&& \\ \cline{1-9}

\multirow{7}{*}{\rotatebox{90}{\textbf{Unidentified}}}&\multirow{7}{*}{\rotatebox{90}{\textbf{Clear}}}&amazonaws.com&\checkmark&\checkmark&&&\checkmark& \\ \cline{3-9}
&&igstudios.in&&&&&&\checkmark \\ \cline{3-9}
&&69.71.216.204&&&\checkmark&&& \\ \cline{3-9}
&&198.105.199.145&&&\checkmark&&& \\ \cline{3-9}
&&akamaitechnologies.com&&&&&&\checkmark \\ \cline{3-9}
&&softlayer.com&&&&&&\checkmark \\ \cline{3-9}
&&cloud-ips.com&&&&&&\checkmark \\ \cline{3-9}
&&mydas.mobi&&\checkmark&&&& \\ \cline{3-9}
&&mkhoj.com&&\checkmark&&&& \\ \cline{3-9}
&&74.217.75.7/8&&&&\checkmark&& \\ \cline{1-9}
\end{tabular}
}
\end{table*}

\begin{table}[h]
\caption{Servers where IdentifierForVendor is communicated in 140 iOS Apps tested.}
\label{table:ServersOfIdForVendor}
\resizebox{\columnwidth}{!}{
\begin{tabular}{|C{3cm}|C{3cm}|C{3cm}|C{3cm}|} 
\hline
\multicolumn{2}{|c|}{\textbf{Third-parties}}&\multicolumn{2}{c|}{\textbf{First-parties}}\\ \hline
\textbf{Clear}&\textbf{SSL}&\textbf{Clear}&\textbf{SSL}\\ \hline \hline
mobileroadie.com,
clara.net,
appads.com,
adcolony.com,
sophiacom.fr,
7mobile7.com,
sitestat.com,
mediatemple.net
&
tapjoyads.com,
tapjoy.com,
adzcore.com,
fiksu.com,
crittercism.com,
ad-x.co.uk
&
eurosport.com,
gameloft.com
&
eamobile.com,
dailymotion.com,
foursquare.com,
google.com,
googleapis.com,
paypal.com \\ \hline
\end{tabular}
}
\end{table}

 \begin{table*}[t]
\caption{User PII transmitted by a total of 140 Android Apps tested.}
\label{table:OtherPrivDataSentToWhere_Android}
\resizebox{2\columnwidth}{!}{
\begin{tabular}{|C{1cm}|C{1cm}||C{3cm}|C{3cm}|C{3cm}|C{3cm}|C{3cm}|C{3.7cm}|C{3.7cm}|}
\cline{3-9}
\multicolumn{2}{c|}{}&\textbf{Server}&\textbf{Accounts}&\textbf{Contacts}&\textbf{Location}&\textbf{Operator Name}&\textbf{SIM Network code}&\textbf{WiFi Scan/Config}\\ \hline \hline

\multirow{21}{*}{\rotatebox{90}{\textbf{Third-parties}}}&\multirow{13}{*}{\rotatebox{90}{\textbf{Clear}}}&seventynine.mobi&&&\checkmark&\checkmark&& \\ \cline{3-9}
&&kiip.me&&&\checkmark&\checkmark&& \\ \cline{3-9}
&&google.com&&&\checkmark&&\checkmark& \\ \cline{3-9}

&&3g.cn&&&\checkmark&&& \\ \cline{3-9}
&&doubleclick.net&&&&&\checkmark& \\ \cline{3-9}
&&goforandroid.com&&&&&\checkmark& \\ \cline{3-9}
&&adtilt.com&&&&\checkmark&& \\ \cline{3-9}
&&2o7.net&&&&\checkmark&& \\ \cline{3-9}
&&nexage.com&&&&&\checkmark& \\ \cline{3-9}
&&ad-market.mobi&&&&&\checkmark& \\ \cline{3-9}
&&mopub.com&&&\checkmark&&& \\ \cline{3-9}
&&mydas.mobi&&&\checkmark&&& \\ \cline{3-9}
&&startappexchange.com&&&&&\checkmark& \\ \cline{2-9}

&\multirow{8}{*}{\rotatebox{90}{\textbf{SSL}}}&airpush.com&&&\checkmark&\checkmark&& \\ \cline{3-9}
&&appwiz.com&&&\checkmark&&\checkmark& \\ \cline{3-9}
&&agoop.net&&&\checkmark&&\checkmark& \\ \cline{3-9}
&&tapjoyads.com&&&&\checkmark&\checkmark& \\ \cline{3-9}

&&crittercism.com&&&&\checkmark&& \\ \cline{3-9}
&&inmobi.com&&&&&&\checkmark \\ \cline{3-9}
&&appsflyer.com&&&&\checkmark&& \\ \cline{3-9}
&&googleapis.com&\checkmark&&&&& \\ \cline{1-9}

\multirow{7}{*}{\rotatebox{90}{\textbf{First-parties}}}&\multirow{2}{*}{\rotatebox{90}{\textbf{Clear}}}&betomorrow.com&&&\checkmark&\checkmark&\checkmark& \\ \cline{3-9}
&&avast.com&\checkmark&&&\checkmark&& \\ \cline{2-9}

&\multirow{5}{*}{\rotatebox{90}{\textbf{SSL}}}&google.com&\checkmark&\checkmark&&&\checkmark& \\ \cline{3-9}
&&badoo.com&&&\checkmark&\checkmark&\checkmark&\checkmark \\ \cline{3-9}

&&checkmin.com&&&\checkmark&&& \\ \cline{3-9}
&&groupon.de&&&&\checkmark&& \\ \cline{3-9}
&&m6replay.fr&&&&\checkmark&& \\ \cline{1-9}

\multirow{5}{*}{\rotatebox{90}{\textbf{Unidentified}}}&\multirow{5}{*}{\rotatebox{90}{\textbf{Clear}}}&91.103.140.193&\checkmark&&&\checkmark&\checkmark& \\ \cline{3-9}
&&adkmob.com&&&&&\checkmark& \\ \cline{3-9}
&&dsxvr.com&&&&&\checkmark& \\ \cline{3-9}
&&amazonaws.com&&&&\checkmark&& \\ \cline{3-9}
&&183.61.112.40&&&&&\checkmark& \\ \cline{1-9}
\end{tabular}
}
\end{table*}

\begin{table*}[t]
\caption{User PII transmitted by a total of 140 iOS Apps tested.}
\label{table:OtherPrivDataSentToWhere_iOS}
\resizebox{2\columnwidth}{!}{
\begin{tabular}{|C{1cm}|C{1cm}||C{3cm}|C{3cm}|C{3cm}|C{3cm}|C{3cm}|C{3.7cm}|C{3cm}|}
\cline{3-9}
\multicolumn{2}{c|}{}&\textbf{Server}&\textbf{Accounts}&\textbf{AddressBook}&\textbf{Device Name}&\textbf{Location}&\textbf{SIM Network Name}&\textbf{SIM Number}\\ \hline \hline
\multirow{8}{*}{\rotatebox{90}{\textbf{Third-parties}}}&\multirow{3}{*}{\rotatebox{90}{\textbf{Clear}}}&clara.net&&&&&\checkmark&\\ \cline{3-9} 
&&amazonaws.com&&&&&\checkmark& \\ \cline{3-9}
&&bkt.mobi&&&&\checkmark&& \\ \cline{2-9}

&\multirow{5}{*}{\rotatebox{90}{\textbf{SSL}}}&capptain.com&&&&\checkmark&\checkmark& \\ \cline{3-9}

&&fring.com&&&&&&\checkmark \\ \cline{3-9}
&&crittercism.com&&&\checkmark&&& \\ \cline{3-9}
&&boxcar.io&&&\checkmark&&& \\ \cline{3-9}
&&testflightapp.com&&&&&\checkmark& \\ \cline{1-9}
\multirow{8}{*}{\rotatebox{90}{\textbf{First-parties}}}&\multirow{3}{*}{\rotatebox{90}{\textbf{Clear}}}&mobilevoip.com&&\checkmark&&&& \\ \cline{3-9}
&&groupon.de&&&&&\checkmark& \\ \cline{3-9}
&&sncf.com&&&&\checkmark&& \\ \cline{2-9}
&\multirow{5}{*}{\rotatebox{90}{\textbf{SSL}}}&groupon.de&&&&\checkmark&\checkmark& \\ \cline{3-9}
&&ebay.com&&&&&&\checkmark \\ \cline{3-9}
&&foursquare.com&&&&\checkmark&& \\ \cline{3-9}
&&paypal.com&&&\checkmark&&& \\ \cline{3-9}
&&twitter.com&\checkmark&&&&& \\ \cline{1-9}
\end{tabular}
}
\end{table*}

\begin{table*}
 \centering
 \caption{Different Pasteboard Names, Types and Values created by 140 iOS Apps tested.}
 \label{table:listOfPBNamesTypesValues}
 \resizebox{2\columnwidth}{!}{
 \begin{tabular}{|C{8cm}|C{8cm}|C{8cm}|} 
 \hline
 \textbf{Pasteboard Names}&\textbf{Pasteboard Types}&\textbf{Pasteboard Values}\\ \hline\hline
 fb\_app\_attribution, org.OpenUDID.slot.0 to 99, com.hasoffers.matsdkref, 
 com.ad4screen.bma4s.dLOG, com.ad4screen.bma4savedata124780,
 com.flurry.pasteboard, com.fiksu.288429040-store, com.fiksu.pb-0 to 19, org.secureudid-0 to 99, 
 com.ebay.identity, com.paypal.dyson.linker\_id, AmazonAdDebugSettings, CWorks.5cb7c5449e677be888147c58,
 amobeePasteboard, com.google.maps,
 com.google.plus.com.deezer.Deezer, com.bmw.a4a.switcher.featureInfos
 and many more
 & com.crittercism.uuid, org.OpenUDID,public.utf8-plain-text, com.fiksu.id, public.secureudid, com.google.maps.SSUC, com.flurry.UID, com.bmw.a4a.switcher.featureinfo container,                     
 &WiFi MAC Address, 2501110D-69B7-415A-896B-4F7A83591263, ID521411E3-D88E-426E-9B7D-1060C0772C89969DC466, 363046414344413130433230, 8211d087-ca5b-42c3-a1a2-7b3779f6c206, 81C65A17-9F0E-4BFE-83A7-1C2C070C3353, E6644EEB-04B3-4AEF-8562-A2C29E323CCE, 55b0a791-517e-4bd4-8398-414dd527417b, And other binary data instances      \\ 
 \hline
 \end{tabular}
 }
 \end{table*}

 \begin{table*}[t]
\centering
\caption{List of third-parties knowing names of installed packages on Android (out of a total of 140 Apps tested).}
\label{table:ThirdPartyWithProcessNamesKnown_Android}
\resizebox{2\columnwidth}{!}{
\begin{tabular}{|C{5cm}|C{18cm}|} 
\hline
\textbf{Third-party (Comm type)}&\textbf{Process Names}\\ \hline\hline
trademob.com(SSL) & All the processes running on the phone
\\ \hline
google.com(SSL) & All the processes running on the phone
\\ \hline
google-analytics.com(SSL)&
com.anydo,
  com.rechild.advancedtaskkiller,
  com.spotify.mobile.android.ui,
  com.google.android.googlequicksearchbox,
  com.dailymotion.dailymotion,
  com.aa.android,
  com.comuto,
  com.airbnb.android \\ \hline
doubleclick.net(plain-text)&
com.tagdroid.android,
  com.rechild.advancedtaskkiller,
  bbc.mobile.news.ww,
  ua.in.android\_wallpapers.spring\_nature \\ \hline
crashlytics.com(SSL)&
com.evernote,
  com.path,
  com.lslk.sleepbot,
  com.twitter.android,
  com.dailymotion.dailymotion \\ \hline

\end{tabular}
}

\end{table*}

\begin{table*}[t]
\centering
\caption{List of third-parties knowing names of installed packages on iOS (out of a total of 140 Apps tested).}
\label{table:ThirdPartyWithProcessNamesKnown}
\resizebox{2\columnwidth}{!}{
\begin{tabular}{|C{4.5cm}|C{21cm}|} 
\hline
\textbf{Third-party (Comm type)}&\textbf{Process Names}\\ \hline \hline
flurry.com(plain-text)&
TopEleven,
  Bible,
  RATP,
  Transilien,
  TripIt,
  DespicableMe,
  FlyAirIndia,
  Viadeo,
  Bankin',
  VDM,
  OCB,
  DuplexA86,
  SleepBot,
  Snapchat,
  Appygraph,
  Booking.com,
  foodspotting,
  Badoo,
  EDF-Releve,
  WorldCup2011,
  Quora,
  UrbanDictionary,
  babbelSpanish,
  MyLittleParis,
  Volkswagen\\ \hline
google-analytics.com(SSL)&
InstantBeautyProduction, Evernote, LILIGO,
  Transilien,
  Viadeo,
  VDM,
  comuto,
  easyjet,
  VintedFR,
  Volkswagen \\ \hline
crashlytics.com(SSL)&
dailymotion,
  TopEleven,
  AmazonFR,
  Path,
  RunKeeper,
  foodspotting,
  babbelSpanish,
  Deezer\\ \hline
urbanairship.com(SSL)&
Wimbledon,
  RATP,
  HootSuite,
  DuplexA86,
  Appygraph,
  foodspotting,
  Volkswagen\\ \hline
  xiti.com(plain-text)&
laposte,
  ARTE,
  myTF1,
  lequipe,
  SoundCloud,
  20minv3,
  Leboncoin\\ \hline
  admob.com(plain-text)&
VSC,
  BBCNews,
  WorldCup2011,
  RF12,
  UrbanDictionary\\ \hline
  capptain.com(plain-text)&
Viadeo,
  myTF1,
  rtl-fr-radios,
  20minv3,
  iDTGV\\ \hline
tapjoy.com(SSL)&
TopEleven,
  Bible,
  DespicableMe,
  OCB,
  MCT\\ \hline
\end{tabular}
}
\end{table*}

\end{document}